\documentclass[final,3p,times]{elsarticle}

\usepackage{booktabs}
\usepackage{bigdelim}
\usepackage{graphicx}
\usepackage[utf8]{inputenc}
\usepackage[T1]{fontenc}
\usepackage[english]{babel}
\usepackage{amssymb}
\usepackage{amsmath}
\usepackage{graphicx}
\usepackage{subcaption}
\usepackage{graphicx}
\usepackage{float}
\usepackage{tabularx}
\usepackage{ulem}
\usepackage{epstopdf}
\usepackage{textcomp}
\usepackage{xcolor}
\usepackage{gensymb}
\usepackage[toc,page,titletoc]{appendix}
\usepackage[capitalise,nameinlink]{cleveref}
\usepackage{multirow}
\usepackage{array}
\newcolumntype{C}[1]{>{\centering\arraybackslash}m{#1}}
\usepackage{makecell}
\usepackage{cleveref}
\setcellgapes{4pt}
\usepackage{xcolor}
\usepackage{setspace}
\usepackage{url}
\usepackage{booktabs}
\usepackage{bigdelim}
\usepackage{graphicx}
\usepackage{array}

\newenvironment{hebtab}{%
  \fontsize{8.0}{9.6}\selectfont
  \setlength{\tabcolsep}{1.76pt}%
  \setlength{\arrayrulewidth}{0.4pt}%
}{}

\newcommand{\hebhead}[1]{{\fontsize{5.76}{6.91}\selectfont\textit{#1}}}

\newcommand{\hebrulebold}[1]{%
  \noalign{\global\arrayrulewidth=1.1pt}\cline{#1}%
  \noalign{\global\arrayrulewidth=0.4pt}}
\newcommand{\hebruleouter}[1]{%
  \noalign{\global\arrayrulewidth=0.9pt}\cline{#1}%
  \noalign{\global\arrayrulewidth=0.4pt}}

\newcommand{\hebsection}[2]{\multicolumn{#1}{|l|}{\textbf{#2}}}

\usepackage{xr}
\usepackage{placeins}
\makeatletter
\newcommand*{\addFileDependency}[1]{%
  \typeout{(#1)}%
  \@addtofilelist{#1}%
  \IfFileExists{#1}{}{\typeout{No file #1.}}%
}
\makeatother
\newcommand*{\myexternaldocument}[1]{%
  \externaldocument{#1}%
  \addFileDependency{#1.tex}%
  \addFileDependency{#1.aux}%
}
\myexternaldocument{SI}
\journal{Nature}

\begin{document}

\begin{frontmatter}

\title{Instability-driven segregation of V group transition metals on HEB$_2$ grain boundaries influencing enhanced mechanical properties}

\author[label1]{Daniil Alekseev\corref{cor1}}
\author[label1]{Viktor S. Baidyshev}
\author[label1]{Alexander G. Kvashnin}

\cortext[cor1]{Corresponding author. E-mail: Daniil.Alekseev@skoltech.ru}

\address[label1]{Skolkovo Institute of Science and Technology, Bolshoy Boulevard 30, bld. 1, Moscow 121205, Russian Federation}

\begin{abstract}
Five metals share a common sublattice in the high-entropy diboride (Hf,Ta,Ti,Nb,Zr)B$_2$ (HEB$_2$).
Most high-entropy materials are produced as polycrystalline materials in the form of sintered ceramics.
The question of which metal species are located at grain boundaries is still unknown. 
Here, we calculated the grain-boundary energy $\gamma_{GB}$ of four types of boundaries in five binary diborides and HEB$_2$ using density functional theory (DFT).
We analyzed the bonding at these boundaries using COHP and sampled cation segregation at six boundaries of HEB$_2$, ordered and general, using Monte Carlo / molecular dynamics with a MACE potential that was fine-tuned for this system.
According to our results $\gamma_{GB}$ splits by group: the group V diborides NbB$_2$ and TaB$_2$ form boundaries with about half the energy of the group IV at three of the four types of GBs, ranging from 0.8 to 2.7 compared to 2.5--4.8~J/m$^2$.
COHP shows a significant difference at the Fermi level, namely the Me--B bonds of NbB$_2$ and TaB$_2$ are antibonding at Fermi energy ($E_F$) in the bulk but bonding at the GB, while their integrated strength hardly changes. 
In HEB$_2$ the segregation follows a similar pattern: Nb and Ta are enriched at five of the six grain boundaries, including general boundaries, up to 23--35~at.\% at 2000~K, although the size of the atoms would favor Zr and Hf in the expanded sites.
Only the $\Sigma 7$ tilt grain boundary, with compressed sites adjacent to expanded ones, sorts according to their size.
These site preferences alone reduce the grain boundary energy $\gamma_{GB}$ by 0.25--0.74~J/m$^2$ at 2000~K.
A segregated $\Sigma 193$ grain boundary can withstand 26 against 20~GPa for a random boundary in tension at 300~K.
The grain boundaries of high-entropy diboride are composed of group V metals, as suggested by the boundary energies of binaries compounds and their bonding at the Fermi level.
Grain boundaries of the high-entropy diboride (Hf,Ta,Ti,Nb,Zr)B$_2$ preferentially include Nb and Ta.
This lowers the grain boundary energy and strengthens the boundaries.
This is consistent with the lower boundary energy and changed Fermi level bonding in NbB$_2$ and TaB$_2$.
\end{abstract}

\end{frontmatter}

\section{Introduction}

Transition-metal diborides MeB$_2$ (Me = Ti, Zr, Hf, Nb, Ta) melt at 3280--3650~K\cite{thorpe2026melting,fahrenholtz2007refractory,magnuson2022review}, have a hardness of about 20--30~GPa for dense material\cite{fahrenholtz2007refractory,cappai2025individual} and resist oxidation up to about 1100~$^\circ$C, while a layer of liquid B$_2$O$_3$ covers the surface\cite{fahrenholtz2007refractory}.
This makes them candidates for the leading edges of hypersonic vehicles and for rocket nozzles, and TiB$_2$ is already used for armor and cutting tools~\cite{fahrenholtz2007refractory}. 
Structure of transition-metal diborides belongs to AlB$_2$-type ($P6/mmm$) containing flat hexagonal boron nets alternate with metal layers, and the strong B--B and Me--B bonds carry the hardness and the stiffness\cite{magnuson2022review}. 
The bonding follows the valence electron count (VEC, here per formula unit with three electrons per boron atom\cite{hossain2026superhard}). 
For the group IV metals (Ti, Zr, Hf) VEC = 10 and the Fermi level falls into the pseudogap between bonding and antibonding states.
The extra electron of the group V metals Nb and Ta (VEC = 11) goes into antibonding states\cite{vajeeston2001electronic,fahrenholtz2007refractory,zhai2026significant}.
It is the same band-filling argument that puts the hardness maximum of the transition-metal carbonitrides at an intermediate electron count\cite{jhi1999electronic}.

Mixing the five metals in equal proportions on the metal sublattice gives the high-entropy diboride (Hf,Ta,Ti,Nb,Zr)B$_2$ (HEB$_2$), which is single-phase after spark plasma sintering at 2000~$^\circ$C\cite{gild2016highentropy}. 
Its hardness is close to the rule of mixtures of the binaries\cite{gild2016highentropy,feng2020processing}, and it oxidizes slower than all of them but HfB$_2$\cite{gild2016highentropy}. 
With VEC = 10.4 it sits just above the optimum, because across a series of high-entropy diborides the calculated hardness is highest at VEC = 10\cite{hossain2026superhard}.

But a sintered ceramic is not a single crystal. 
Diborides densify only at temperatures where grain-boundary and volume diffusion become fast\cite{fahrenholtz2007refractory}, and their strength and hardness depend on the grain size. 
The strength of ZrB$_2$ scales roughly with the inverse square root of the grain size\cite{fahrenholtz2007refractory}, the hardness of a high-entropy diboride follows the Hall--Petch law\cite{yang2024effect}, and adding Nb to (Hf,Zr,Ti,Ta)B$_2$, where it forms a second, Nb-rich boride, refines the grains and raises the hardness\cite{feng2021effect}. 
In HEB$_2$ the grains grow five times slower than in ZrB$_2$, which was attributed to sluggish diffusion and to a boundary energy lowered by segregation, neither of them measured\cite{cheng2023grain}. 
So what a high-entropy boundary is made of, and what it costs, enters both the microstructure and the mechanical response of the ceramic.

DFT gives the grain-boundary energy $\gamma_{GB}$ of a few tilt boundaries in TiB$_2$\cite{dai2020strategy} and ZrB$_2$\cite{dai2017segregation}, and the structure of several [0001] tilt boundaries in both\cite{guo2021superhardness}.
For HfB$_2$, NbB$_2$ and TaB$_2$ there are no available information.
Segregation was computed for dilute solutes in these hosts, and there it mostly follows size, where large solutes go to expanded boundary sites and small ones to compressed sites\cite{dai2017segregation,dai2020strategy}, which in ZrB$_2$ sends Nb and Ta to the compressed ones\cite{dai2017segregation}.
At the expanded $\Sigma 6$ site of TiB$_2$, though, Nb and Ta segregate more strongly than the larger Zr and Hf\cite{dai2020strategy}.

Machine-learning potentials obtained the same picture at finite temperature, where W segregated to the boundaries of ZrB$_2$ and strengthened nearly all of them\cite{dai2020grain}, and in (Ti,Zr,Hf)B$_2$ the smallest cation, Ti, went to the boundaries\cite{dai2024deep}. 
In related five-metal diborides, experiments found W and Mo replacing some of the host metals\cite{wang2021synergic}, and Er replacing Nb\cite{qin2022processingdependent} at the boundaries.
Next to Er, Zr and Hf were depleted, and Ta was not enriched\cite{qin2022processingdependent}. 
In HEB$_2$ itself, the five metals looked homogeneous across a boundary in STEM-EDS maps\cite{wang2021synergic}. 
For the five-metal diboride there was no $\gamma_{GB}$ and no calculated segregation.

The electron count suggests a possible answer. 
At the boundary the boron nets reconstruct and the Me--B and Me--Me bonds are reformed with different lengths and angles.
If the extra electron of group V is in the antibonding states of these bonds in the bulk, the reconstruction that moves these states away from the Fermi level is less costly for Nb and Ta than for Ti, Zr and Hf.
This should cause a splitting of $\gamma_{GB}$. 
In HEB$_2$, where both groups share one sublattice, metals with cheaper boundaries should segregate.
As Nb, with the lowest boundary energy, does in the NbMoTaW alloy\cite{li2020complex} (where Ta, also from group V, is depleted), this would put Nb and Ta on the boundaries even at expanded sites.
This is because size alone would send Zr and Hf there.
Being a hypothesis about electron count, this can be tested on five binaries and then checked in a mixture.
Since a COHP analysis has already supported a bonding origin of solute segregation to a boundary in Fe\cite{ito2021electronic}, the same analysis can be done at the boundary itself.

DFT with COHP provides $\gamma_{GB}$ and bonding at a few boundaries, in cells containing a few hundred atoms, with one decoration of the metal sublattice at a time.
Segregation requires thousands of atoms, thermal disorder and many decorations, which DFT cannot accommodate.
The deep potential available for nine diborides\cite{dai2024deep} has no grain boundaries in its training set, and its authors found it less accurate for boundary energies than for other properties
Therefore, we built our own MACE model\cite{mace,macemh1} fine-tuned on a dataset assembled using active learning\cite{activemiao} and complemented by hand with amorphous and surface configurations that sample disordered and under-coordinated configurations of the boundary core.
Boundaries were excluded from training, and the potential reproduced DFT $\gamma_{GB}$ sufficiently well to rank boundaries and metals.

Here we first compute, in DFT, $\gamma_{GB}$ of four types of grain boundaries ($\Sigma 6$, $\Sigma 7$ tilt, $\Sigma 7$ twist, $\Sigma 8$) in the five binary diborides and in HEB$_2$ modeled by special quasirandom structures (SQS)\cite{sqs}.
Then we compare the bonding at the boundary with the bulk by using COHP, and then sample cation segregation at six boundaries of HEB$_2$, ordered and general, by hybrid Monte Carlo / molecular dynamics.
The $\gamma_{GB}$ value of the group V diborides is approximately half that of the group IV at three out of the four boundaries. 
The COHP reveals the difference at the Fermi level.
The Me--B bonds in NbB$_2$ and TaB$_2$ are antibonding at Fermi energy ($E_F$) in the bulk and bonding at the boundary. 
In HEB$_2$ Nb and Ta segregate to five out of the six boundaries, including general boundaries.
Only at the $\Sigma 7$ tilt, which offers compressed sites next to expanded ones, does size prevail, and Ti and Zr segregate by radius.
Finally, we pull the single crystal and the $\Sigma 193$ boundary to failure.
The segregated boundary is stronger than a random boundary by 31\% (25.9 against 19.7~GPa at 300~K).

\section{Methods}

\subsection{DFT calculations}
\label{sec:dft}

DFT calculations are performed in VASP\cite{vasp} using the Perdew--Burke--Ernzerhof (PBE) functional\cite{pbe} and Ti\_sv, Zr\_sv, Nb\_sv, Hf\_pv, Ta\_pv and B PAW datasets\cite{paw,paw_kresse}.
The kinetic energy cutoff is set to 500 eV with $k$-point mesh resolution of 0.16~\AA$^{-1}$ and Gaussian smearing of width $\sigma = 0.05$ eV.
The convergence criterion for energy for the electronic self-consistent field (SCF) loop is set to $10^{-6}$~eV.
PBE was chosen against RPBE\cite{rpbe} and PBEsol\cite{pbesol} by comparison with r2SCAN\cite{r2scan} on NbB$_2$, where it gave the closets results on the (0001) surface energy (Supplementary Sec.~\ref{si:dft-xc}). NbB$_2$ was chosen for convergence tests because Nb exhibits a pronounced sensitivity of the band energy to Fermi-level shifts\cite{du2026origins}, which makes total energies and forces in NbB$_2$ particularly sensitive to Brillouin-zone sampling and smearing. So we consider that settings converged for NbB$_2$ are therefore conservative for the other diborides.
Set parameters for energy cutoff and $k$-point mesh resolution give are converged to about 1~meV/atom in absolute energy and better in energy differences (Supplementary Sec.~\ref{si:dft-conv}). 
Grain-boundary cells are relaxed with respect to atoms, cell shape and volume are relaxed down to forces of 0.01~eV/\AA{} at $k$-point mesh resolution of 0.30~\AA$^{-1}$, which fits into the memory of a single V100S GPU, and then recomputed as single points at 0.16~\AA$^{-1}$.
A relaxation at 0.16~\AA$^{-1}$ changes $\gamma_{GB}$ by at most 0.004~J/m$^2$ (Supplementary Sec.~\ref{si:dft-kspacing}). 
Slabs are relaxed with respect to atoms only, in the cell of the relaxed boundary.

Grain boundary energy $\gamma_{GB}$ (excess energy of GB compared to bulk per surface area), surface energy $\gamma_{s}$ (excess energy of free surface compared to bulk per surface ares) and work of separation $W_{sep}$ (energy it takes to split the GB into two different surfaces) of a configuration with area $S$ parallel to the boundary or surface are calculated as:

\[
\gamma_{GB} = \frac{E_{GB} - E_{bulk}}{2S}, \qquad \gamma_{s} = \frac{E_{s} - E_{bulk}}{2S}, \qquad W_{sep} = 2 \gamma_s - \gamma_{GB}
\]

, where $E$ is the total energy of the considered system (grain boundary, surface, bulk). 
For HEB$_2$ the boundary cells are equimolar (Sec.~\ref{sec:gbc}), and $E_{bulk}$ is their number of atoms times the energy per atom of an equimolar bulk SQS cell (90 atoms, Ti$_6$Zr$_6$Nb$_6$Hf$_6$Ta$_6$B$_{60}$, mean over three decorations).
% verify: equimolar bulk reference pending (homer job 1320, sqs1-3); Table 1 HEB2 column, Sec. 3.1 numbers and the HEB2 test points of Fig. 2 / Table 2 still use the old 54-atom shifted reference

\subsection{Construction of grain boundaries and slabs}
\label{sec:gbc}

Four boundaries are considered in DFT: the
$\Sigma 7\,[0001]$ $21.8^\circ$ symmetric tilt boundary on the
$(12\bar{3}0)$ plane, the $\Sigma 7\,[0001]$ $21.8^\circ$ twist boundary
on the basal plane, a coherent $(01\bar{1}1)$ twin boundary
($78^\circ$ rotation about $[2\bar{1}\bar{1}0]$, conventionally labeled
$\Sigma 8$), and an asymmetric
$(\bar{2}11\bar{7})\,\|\,(\bar{1}2\bar{1}\bar{1})$ boundary
($47.6^\circ$ rotation about $[10\bar{1}0]$, labeled $\Sigma 6$). 
For $\Sigma 6$ and $\Sigma 8$ the rotation axis lies in the basal plane, where exact coincidence needs a rational $c^2/a^2$, so their $\Sigma$ labels are nominal (construction and coincidence mismatch in Supplementary Sec.~\ref{si:gb-dft}). 
The cells hold 105 ($\Sigma 6$), 78 ($\Sigma 7$ tilt), 48 ($\Sigma 8$) and 210 ($\Sigma 7$ twist) atoms, two boundaries each, and are shown in the top row of Fig.~\ref{fig:gb_parity} for TiB$_2$. 
For HEB$_2$ all four grain boundaries are computed in equimolar SQS cells: $\Sigma 6$ in the 105-atom cell, relaxed in DFT, and the twist, the $\Sigma 7$ tilt and $\Sigma 8$ in cells of 210, 390 and 240 atoms, which are relaxed with the MACE potential of Sec.~\ref{sec:potential}, since a DFT run of an HEB$_2$ cell takes much longer than one of a binary, and then computed as DFT single points.
The slabs are cleaves of the boundary cells, so for each $\gamma_{GB}$ the corresponding $\gamma_{s}$ is used.

For MC/MD the four grain boundaries are rebuilt in cells of 1584--2304 atoms, together with one more ordered boundary, a $(05\bar{5}2)$ twin ($35.56^{\circ}$ rotation about $[2\bar{1}\bar{1}0]$, labeled $\Sigma 193$, nominal in the same sense as $\Sigma 6$ and $\Sigma 8$) in a cell of 2340 atoms, and a three-grain polycrystal of 2919 atoms with random grain orientations, generated with Atomsk\cite{atomsk} (Supplementary Sec.~\ref{si:gb-mlip}).

\subsection{Preparation of the machine-learning potential}
\label{sec:potential}

\begin{figure}[H]
    \centering
    \includegraphics[width=\linewidth]{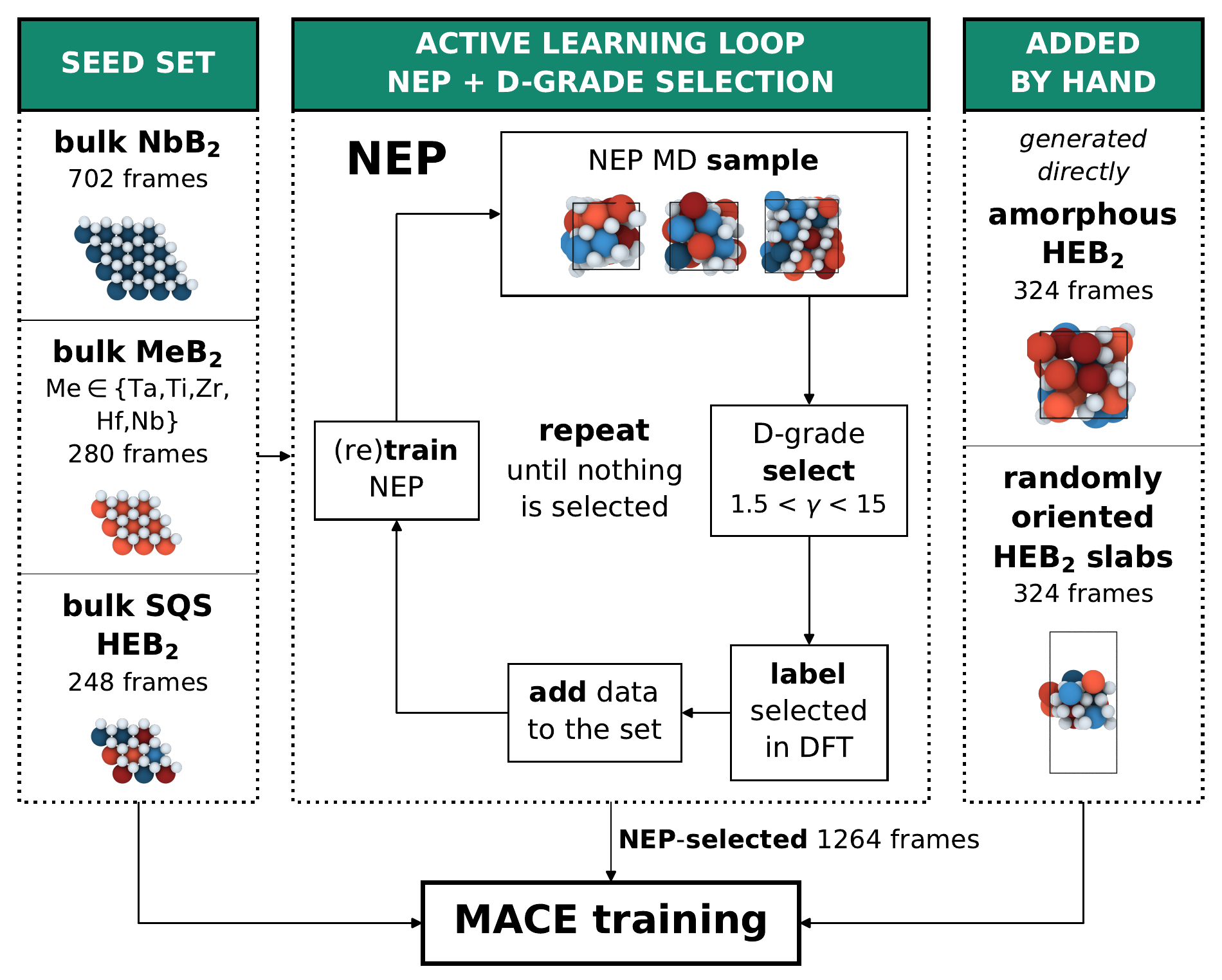}
    \caption{Assembly of the training set. Left: seed set of bulk NbB$_2$ (702 frames), the five binar diborides MeB$_2$, Me $\in$ \{Ta, Ti, Zr, Hf, Nb\} (280 frames) and SQS HEB$_2$ (248 frames). Middle: active-learning loop, in which a NEP potential is (re)trained, MD with this NEP generates candidate configurations, the candidates are screened by the extrapolation grade $1.5 < \gamma < 15$ (D-optimality), the selected ones are labeled in DFT and added to the set; the loop is repeated until nothing passes the selection step. Right: amorphous HEB$_2$ (324 frames) and randomly oriented HEB$_2$ slabs (324 frames) added by hand. The production potential (MACE) is fine-tuned on the union of all three blocks.}
    \label{fig:scheme}
\end{figure}

The training set is built in three blocks, shown in Fig.~\ref{fig:scheme}.
Each block is labeled with the DFT protocol described in Sec.~\ref{sec:dft}.
The seed set comprises the five binary diborides, strained and heated special quasirandom structure (SQS) cells of HEB$_2$, and the corresponding elastically deformed configurations. NbB$_2$ is overrepresented in the seed set (highlighted here and in Fig.~\ref{fig:scheme}) because it served as the test system for the methodological experiments, including the convergence tests. It was chosen for this role as the most demanding case among the binary diborides: Nb is one of the elements on which machine-learning interatomic potentials consistently show elevated force errors, a difficulty attributed to the strong response of its electronic band energy to Fermi-level shifts~\cite{du2026origins}.
The configurations that were not included in the initial set were collected using active learning (AL) with a NEP4 potential\cite{nep} in GPUMD\cite{gpumd} with implemented D-optimality framework\cite{activemiao} which uses the extrapolation grade $\gamma$ of a local environment computed by MaxVol over the NEP descriptors.
The same construction is implemented in the MLIP code for active learning of Moment Tensor Potentials \cite{mtp,mlip,mlip3}.
Configurations of NEP MD with $1.5 < \gamma < 15$ are labeled and added until none is left, see middle part in the Fig.\ref{fig:scheme}.
A configuration for which the extrapolation grade is below 1.5 is considered as interpolation, if the grade is above 15 the configuration is considered as unphysical. 
90 iterations of AL is provided, see Supplementary Sec.~\ref{si:al}. 
The loop samples only NEP MD visits, so 324 amorphous HEB$_2$ cells and 324 randomly oriented HEB$_2$ slabs are added manually for the disordered and under-coordinated environments of the boundary core, see right part in the Fig.\ref{fig:scheme}. 
The final dataset contains 3142 training configurations and 110 validation configurations (Supplementary Table~\ref{tab:dataset})
No grain boundaries are included in the dataset.

The production potential is a fine-tune of the MACE-MH-1 foundation model\cite{mace,macemh1} on the entire dataset. 
The NEP of the loop was not used further, as its energy offset drifted over the AL iterations.
On grain boundary energies it is 2--3 times less accurate than MACE (Supplementary Sec.~\ref{si:nep-mace}). 
The fine-tuning settings and comparison with three other foundation models can be found in Supplementary Secs.~\ref{si:arms} and~\ref{si:finetune}.
The accuracy results are presented in Sec.~\ref{sec:res-pot}.

\subsection{Molecular Dynamics simulations}
\label{sec:md}

Molecular dynamics simulations are carried out in LAMMPS\cite{lammps} with the MACE potential.
Without domain decomposition one V100S GPU holds about 3600 atoms, which sets the cell sizes. 
The cation sublattice of each cell is decorated by annealing the Warren--Cowley parameters\cite{warrencowley} of the first metal shells to zero.
Segregation is sampled by hybrid Monte Carlo/molecular dynamics MC/MD at 2000, 2500 and 3000~K using NPT MD at zero pressure, 120~ps per run.
Metropolis swaps are used to exchange the metal species every 100 timesteps, with boron never being swapped. 
Boundary cation sites are found from the boron coordination number of each metal atom, $CN_B$ for short, as number of B atoms within 3.5 \AA; 12 in the bulk, or for $\Sigma 6$ and $\Sigma 8$ from the shape of its boron cage.
Run settings, site definitions and diffusion coefficients are provided in Supplementary Secs.~\ref{si:md-runs}, \ref{si:md-sites} and~\ref{si:md-diff}.

Segregation energies are calculated in two ways.
The first method is the swap spectrum of a relaxed 0~K cell with random decoration.
In this method every boundary site $s$ is exchanged with each of the four other species taken from a bulk site $b$, and then a local relaxation is performed,

\[
\Delta E(s, A \to i) = E(i \text{ at } s,\ A \text{ at } b) - E(A \text{ at } s,\ i \text{ at } b),
\]
where $A$ is the species that occupied $s$. 
Per site the five energies are shifted to zero mean, $E_i(s) = \Delta E(s, A \to i) - \langle \Delta E(s, A \to j) \rangle_j$ with $\Delta E(s, A \to A) = 0$, so that they do not depend on which cation sits at the site, they are reduced to one number per species by the Boltzmann average:

\[
E^{seg}_i = -kT \ln \langle e^{-E_i(s)/kT} \rangle_{s \in GB} + kT \ln \langle e^{-E_i(s)/kT} \rangle_{s \in \mathrm{int}},
\]
at $T = 2000$~K.

The second way is the McLean isotherm\cite{mclean} inverted from the MC/MD compositions.
A cation can only reach a boundary site by exchanging with another cation.
Therefore the isotherm for inversion is the exchange form, which only gives differences in segregation free energies. With the gauge $\sum_i \Delta G_i = 0$:

\[
\Delta G_i = -kT \left( \ln e_i - \langle \ln e_j \rangle_j \right), \qquad e_i = x_i^{GB} / x_i^{\mathrm{int}},
\]
where $x_i^{GB}$ and $x_i^{\mathrm{int}}$ are the fractions of species $i$ at the boundary and in the interior. 
In both routes a negative energy means that the boundary takes the species.
The two routes are not the same quantity: the spectrum is the energy of one exchange in a random boundary at 0~K, the isotherm is the free energy of an equilibrated boundary at $T$, with the interactions between segregants and the vibrational part is included.
The site sets, the outlier cut of the spectrum and the gauge at the $\Sigma 7$ tilt are given in Supplementary Sec.~\ref{si:md-seg}.

$\gamma_{GB}$ after segregation is computed on MC/MD snapshots quenched to 0~K and is compared with $\gamma_{rand}$ of the relaxed initial cell, which carries a random decoration. 
The expected drop in the swap spectrum, $\Delta\gamma(T)$, is the free energy that the boundary sites gain when their occupations are in equilibrium with the bulk rather than random, with site energies $E_i(s)$ (Supplementary Sec.~\ref{si:md-gamma}). 

Uniaxial tension is simulated at a strain rate of $5\times10^{8}$~s$^{-1}$ on a 432-atom HEB$_2$ single crystal and on the $\Sigma 193$ cell
The latter is performed in three segregated cells that are quenched from a 2000~K run and in the same cells with cations permuted randomly, so that only the arrangement of the cations differed within such a pair (Supplementary Sec.~\ref{si:md-tension}). 
Atomic configurations are rendered with OVITO\cite{ovito}.

\subsection{Bond characterization using the COHP analysis}
\label{sec:cohp-methods}

Bonding in bulk diborides and at grain boundaries is compared using the projected COHP method\cite{Dronskowski1993,Deringer2011} in LOBSTER 5.1.1\cite{Maintz2016,Nelson2020} and, for the 168-atom twist cells (which give the same $\gamma_{GB}$ as the 210-atom ones), in gpucohp\cite{gpucohp}, which is our GPU implementation of the same projection scheme, checked against LOBSTER in Supplementary Sec.~\ref{si:cohp}. 
Every relaxed structure is recalculated as a single point with the settings described in Sec.~\ref{sec:dft}, and with symmetry turned off.
The number of virtual energy bands is set to be roughly twice the number of occupied bands
The PAW wavefunctions are then projected onto the pbeVaspFit2015 Slater-type basis with the semicore states of the metals included.
The absolute charge spilling is 0.62--1.71\%. 
Bond strength is compared through the $-$ICOHP integrated up to the Fermi level:

\[
\mathrm{ICOHP}_{ij} = \int_{-\infty}^{E_F} \mathrm{COHP}_{ij}(E)\, dE,
\]
where a pair of atoms $i$, $j$ enters the analysis as a first-shell bond if the distance between them is shorter than 1.25 times the average bond length of the same type (B--B, Me--B or Me--Me) in the bulk diboride.
A boundary atom is an atom that is within 1.2~\AA{} of the boundary plane, as determined by the excess of its per-atom energy.
The boundary atoms are selected once per boundary in the TiB$_2$ cell and transferred to the other compositions through one-to-one site matching.
We use $-$ICOHP only as a relative measure to compare Me--B and Me--Me bonds between the five metals and between the bulk and grain boundaries, not as an absolute bond energy.
Another descriptor is $-$COHP$_{ij}(E_F)$, the value of the curve at the Fermi level, which indicates whether the states of that bond at $E_F$ are bonding or antibonding.
Its dependence on the energy grid, the integration method and for the HEB$_2$ twist, on the MACE relaxation of the cell is checked in Supplementary Sec.~\ref{si:cohp}.

\section{Results and Discussion}

\subsection{Grain-boundary and surface energetics from DFT}
\label{sec:gb-dft}

DFT values of $\gamma_{GB}$ for the four types of grain boundaries, five binary diborides and the HEB$_2$ SQS cells are collected in Table~\ref{tab:gb_dft}.
The surface energies and the works of separation of the same cells, which serve below only as a test of the potential, are given in Supplementary Table~\ref{si:tab-surf} and Fig.~\ref{si:fig-gamma}.

\begin{table}[H]
\centering
\caption{DFT grain-boundary energies $\gamma_{GB}$ of four grain boundaries in the five binary diborides and in one SQS decoration of HEB$_2$, in J/m$^2$. For HEB$_2$, $\Sigma 7$ tilt and $\Sigma 8$ are the equimolar cells of 390 and 240 atoms. These and the HEB$_2$ twist are DFT single points on MACE-relaxed cells.}
\label{tab:gb_dft}
\begin{hebtab}
\begin{tabular}{@{}|l|r|r|r|r|r|r|@{}}
\hebruleouter{1-7}
\textbf{Boundary} & \multicolumn{3}{c|}{\textbf{Group IV}} & \multicolumn{2}{c|}{\textbf{Group V}} & \\
\cline{2-6}
 & \hebhead{TiB$_2$} & \hebhead{ZrB$_2$} & \hebhead{HfB$_2$} & \hebhead{NbB$_2$} & \hebhead{TaB$_2$} & \hebhead{HEB$_2$} \\
\hebrulebold{1-7}
$\Sigma 6$, asymmetric & 4.78 & 4.25 & 4.51 & 2.73 & 2.18 & 3.65 \\
\cline{1-7}
$\Sigma 7$ tilt, symmetric & 2.65 & 2.58 & 2.74 & 2.36 & 2.26 & 2.56 \\
\cline{1-7}
$\Sigma 8$, coherent twin & 2.87 & 2.53 & 2.64 & 1.37 & 0.83 & 2.28 \\
\cline{1-7}
$\Sigma 7$ twist, basal & 2.80 & 2.59 & 2.66 & 1.27 & 0.84 & 1.95 \\
\hebruleouter{1-7}
\end{tabular}
\end{hebtab}
\end{table}

Table~\ref{tab:gb_dft} splits by group. 
For the $\Sigma 6$, $\Sigma 8$ and twist boundaries of the group IV diborides give $\gamma_{GB}$ of 2.5--4.8 J/m$^2$.
The group V diborides gives $\gamma_{GB}$ of 0.8--2.7 J/m$^2$, i.e. roughly two times lower, with TaB$_2$ always the lowest. 
The only grain boundary where the groups do not separate is the symmetric $\Sigma 7$ tilt with the $\gamma_{GB}$ ranging from 2.26 to 2.74 J/m$^2$ for all five metals, a spread of 0.5 J/m$^2$. 
Within a group the order of the metals holds at three out of four grain boundaries (Ti $>$ Hf $>$ Zr at $\Sigma 6$, $\Sigma 8$ and the twist, Hf $>$ Ti $>$ Zr at the tilt; Nb $>$ Ta everywhere), and at the three grain boundaries that split the differences inside a group are 0.2--0.6~J/m$^2$ against 1.2--2.6~J/m$^2$ between the groups.
The $\Sigma 7$ tilt is computed before on the same plane, and our values agree with the earlier DFT results, namely 2.65 against 2.64~J/m$^2$ for TiB$_2$\cite{dai2020strategy} and 2.58 against 2.52~J/m$^2$ for ZrB$_2$\cite{dai2017segregation} (reference values converted from meV/\AA$^2$). 
Previously studied $\Sigma 6$\cite{dai2020strategy,dai2017segregation} is a symmetric tilt about $[01\bar{1}0]$, which differs from our asymmetric one, and their $\Sigma 8$ is the same twin but off-stoichiometric, and its $\gamma_{GB}$ depends on the chemical potential of boron (0.40--1.44~J/m$^2$ in TiB$_2$, 1.35--2.48~J/m$^2$ in ZrB$_2$), so we do not compare it with our stoichiometric twin.
The group split goes with the extra valence electron of Nb and Ta.
At the boundary the Me--B bonds have to reconstruct, and the COHP analysis of Sec.~\ref{sec:cohp} looks for the trace of that electron in the bonding.

For the six-element cells, with one SQS decoration per boundary, HEB$_2$ gives $\gamma_{GB} = 3.65$~J/m$^2$ for $\Sigma 6$ against a average across five binary diborides is 3.69~J/m$^2$, 2.56 against 2.52~J/m$^2$ for the $\Sigma 7$ tilt, 1.95 against 2.03~J/m$^2$ for the twist and 2.28 against 2.05~J/m$^2$ for $\Sigma 8$.
The second, smaller SQS cells of the tilt and of $\Sigma 8$ (78 and 48 atoms, not equimolar, Fig.~\ref{fig:gb_parity}) show how much of this belongs to the decoration.
They give $\gamma_{GB}$ of 2.36 and 2.60~J/m$^2$, 0.2--0.3~J/m$^2$ away from the equimolar cells. 
Within this scatter, the random HEB$_2$ boundary cost the mean of the binary diborides at all four considered grain boundaries.
Chemical disorder by itself does not move $\gamma_{GB}$ by more than this.
The random decorations of the much larger MACE cells of Sec.~\ref{sec:mlip-gb} gives $\gamma_{GB}$ of 3.76, 2.44, 2.02 and 2.24~J/m$^2$, which is in the same order (Fig.~\ref{fig:gb_kinetics}(b)), and are within 0.12~J/m$^2$ of the DFT values, which is below the error of the potential on $\gamma_{GB}$ (Sec.~\ref{sec:res-pot}). 
The same comparison applies to the closest pair of boundaries in the DFT set.
The two grain boundaries of the 105-atom $\Sigma 6$ cell are only 8.4--9.0~\AA{} apart, compared to 31~\AA{} in the MACE cell.
The difference between these two values is 0.11~J/m$^2$, so their interaction within the DFT cell is within the accuracy of the potential.

\subsection{Accuracy of the potential on grain boundaries and surfaces}
\label{sec:res-pot}

\begin{figure}[H]
    \centering
    \includegraphics[width=\textwidth]{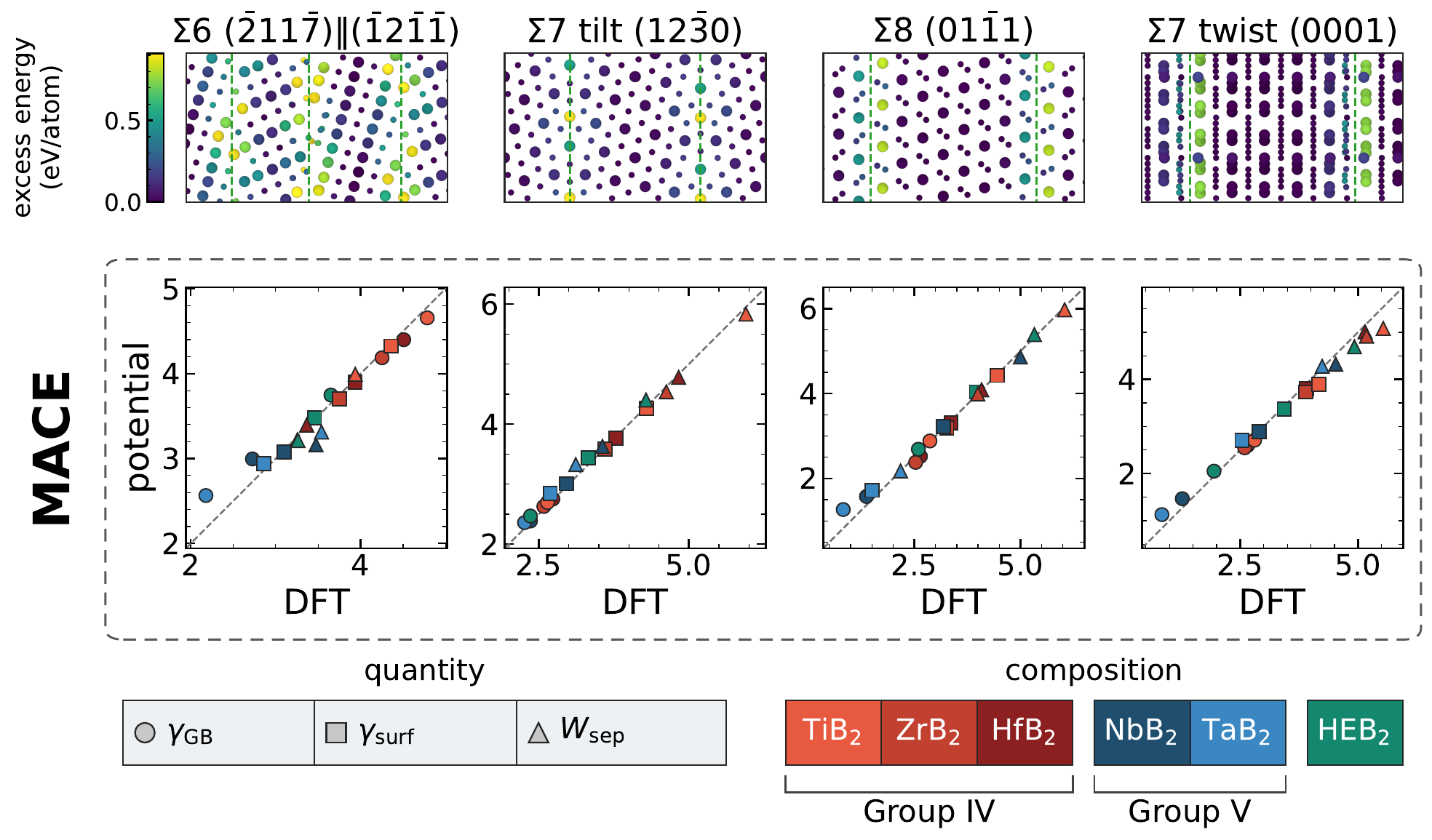}
    \caption{The four grain boundaries of the DFT dataset and the accuracy of MACE on them. Top: relaxed cells (TiB$_2$ is shown) colored by the excess energy per atom with respect to bulk, boundary planes are shown by dashed lines. The boundary plane is given above each panel. Bottom: $\gamma_{GB}$ (circles), $\gamma_{surf}$ of the cleavage plane (squares) and $W_{sep}$ (triangles) predicted by MACE against DFT, one panel per boundary, in J/m$^2$. Colors mark the composition: group IV binaries in red, group V in blue, the SQS HEB$_2$ cell in green (for $\Sigma 7$ tilt and $\Sigma 8$ the smaller SQS cells of 78 and 48 atoms, the same cell in DFT and in the potential, Supplementary Sec.~\ref{si:gb-dft}). Dashed line is $y = x$. The same plot for the NEP of the active-learning loop is Supplementary Fig.~\ref{si:fig-parity-nep}.}
    \label{fig:gb_parity}
\end{figure}

No grain boundary is present in the training set. 
Therefore, the boundaries described in the Table \ref{tab:gb_dft} and their cleavage surfaces are testing the potential of a configuration that the model has not encountered (Fig. \ref{fig:gb_parity}, Table \ref{tab:mlip-benchmark}c).
The surfaces are trained as randomly oriented HEB$_2$ slabs, so the test is weaker for them.
Over the 24 boundary cells (four boundaries in six compositions), the RMSE of $\gamma_{GB}$ is 0.17~J/m$^2$, about 4\% of the 0.8-4.8~J/m$^2$ range.
$W_{sep}$ is reproduced to 0.17~J/m$^2$ and the $(10\bar{1}0)$ surfaces to 0.16~J/m$^2$.
The order of the groups and the metals within each group from Table~\ref{tab:gb_dft} is preserved.
On the six-element cells (green in Fig.~\ref{fig:gb_parity}), MACE stays within 0.02-0.23~J/m$^2$ of DFT for all twelve quantities. 
The same level as for binary diborides.
One systematic error is the compression of the $\gamma_{GB}$ range: the low-energy Nb and Ta boundaries come out 0.2-0.45~J/m$^2$ above DFT at $\Sigma 6$, $\Sigma 8$ and the twist (within 0.1~J/m$^2$ at the tilt). 
So, the MLIP averages are biased against low boundary energies.
The potential is good enough to determine the boundaries and metals, but it is not able to distinguish differences below 0.2~J/m$^2$ between individual grain boundaries.
The errors in the bulk configurations of the training data (2.2~meV/atom and 23~meV/\AA) and in the elastic moduli, and the NEP of the active-learning loop, 2--3 times worse on $\gamma_{GB}$, discussed in Supplementary Section~\ref{si:accuracy}.

\begin{table}[H]
\centering
\caption{RMSE of MACE against DFT. Sections a and b represent validation data, with different configuration types present during training. Section c is the test set, where grain boundaries are not included in the training, and the surfaces are represented only by randomly oriented HEB$_2$ slabs. The NEP of the active-learning loop on the same sets is given in Supplementary Table~\ref{si:tab-benchmark}. $E$ (meV/at) is energy per atom RMSE, $F$ (meV/\AA) is the force per atom RMSE, $\sigma$ (GPa) is stress of the configuration calculated from virial. Elastic moduli $C_{ij}$ in (b) are calculated from finite deformation methods along specific axes. Bulk modulus $B_0$ is calculated from finite deformation along three main axes.}
\label{tab:mlip-benchmark}
\begin{hebtab}
\begin{tabular}{@{}|l|r|r|r|@{}}
\hebruleouter{1-4}
\hebsection{4}{a.~Validation} \\
\cline{1-4}
 & \hebhead{$E$ (meV/at)} & \hebhead{$F$ (meV/\AA)} & \hebhead{$\sigma$ (GPa)} \\
\cline{1-4}
1. Relaxed decorations ($n\!=\!35$) & 2.26 & 9.9 & 0.143 \\
\cline{1-4}
2. Equation of state, $\pm$2--6\% ($n\!=\!60$) & 2.42 & 14.7 & 0.480 \\
\cline{1-4}
3. Uniaxial strain, $\pm$2--8\% ($n\!=\!70$) & 2.26 & 8.7 & 0.192 \\
\cline{1-4}
4. Shear strain, 2--6\% ($n\!=\!48$) & 1.97 & 10.1 & 0.170 \\
\cline{1-4}
5. Size check, 54-atom cells ($n\!=\!5$) & 2.67 & 12.6 & 0.215 \\
\cline{1-4}
6. Thermal MD, 300--2300\,K ($n\!=\!72$) & 2.29 & 42.6 & 0.196 \\
\cline{1-4}
7. Coupled strain\,$+$\,MD ($n\!=\!16$) & 1.77 & 25.5 & 0.184 \\
\cline{1-4}
8. Amorphous ($n\!=\!36$) & 11.29 & 191.0 & 0.750 \\
\cline{1-4}
9. Generated surfaces ($n\!=\!36$) & 8.28 & 174.9 & 0.164 \\
\cline{1-4}
All bulk subsets (1--7) ($n\!=\!306$) & 2.24 & 23.1 & 0.268 \\
\hebrulebold{1-4}
\hebsection{4}{b.~Elastic moduli} \\
\cline{1-4}
 & \hebhead{$B_0$ (GPa)} & \hebhead{$C_{11}$ (GPa)} & \hebhead{$C_{33}$ (GPa)} \\
\cline{1-4}
1. Bulk and axial ($n\!=\!10$) & 1.1 & 6.6 & 14.2 \\
\cline{1-4}
 & \hebhead{$C_{44}$ (GPa)} & \hebhead{$C_{66}$ (GPa)} & \\
\cline{1-4}
2. Shear ($n\!=\!8$) & 9.4 & 11.2 & \\
\hebrulebold{1-4}
\hebsection{4}{c.~Test} \\
\cline{1-4}
 & \hebhead{$\gamma_{\rm GB}$ (J/m$^2$)} & \hebhead{$\gamma_{\rm surf}$, clv.\ (J/m$^2$)} & \hebhead{$W_{\rm sep}$ (J/m$^2$)} \\
\cline{1-4}
1. $\Sigma$6, asymmetric ($n\!=\!6$) & 0.209 & 0.046 & 0.162 \\
\cline{1-4}
2. $\Sigma$7 tilt, symmetric ($n\!=\!6$) & 0.065 & 0.081 & 0.119 \\
\cline{1-4}
3. $\Sigma$8, coherent twin ($n\!=\!6$) & 0.217 & 0.105 & 0.067 \\
\cline{1-4}
4. $\Sigma$7 twist, basal ($n\!=\!6$) & 0.158 & 0.153 & 0.256 \\
\cline{1-4}
All boundaries (1--4) ($n\!=\!24$) & 0.173 & 0.104 & 0.166 \\
\cline{1-4}
 & \multicolumn{3}{c|}{\hebhead{$\gamma_{\rm surf}(10\bar 10)$ (J/m$^2$)}} \\
\cline{1-4}
5. Free surfaces ($n\!=\!5$) & \multicolumn{3}{c|}{0.156} \\
\hebruleouter{1-4}
\end{tabular}
\end{hebtab}
\end{table}

\subsection{HEB$_2$ grain boundary segregation inferred with MACE}
\label{sec:mlip-gb}

\begin{figure}[H]
\centering
\includegraphics[width=\textwidth]{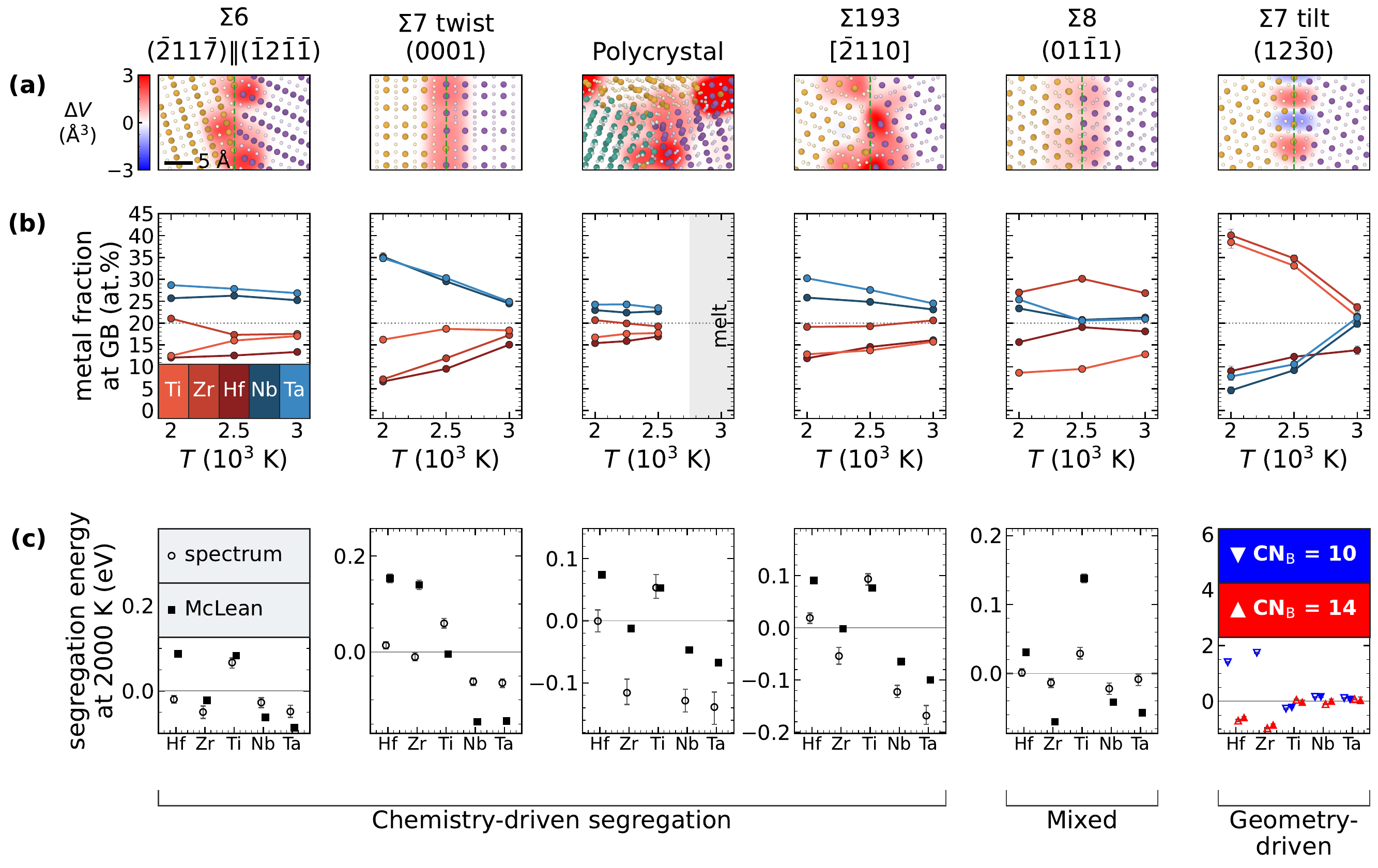}
\caption{Cation segregation at the six grain boundaries of HEB$_2$ from MC/MD simulations with MACE.
(a)~Site volume $\Delta V$ of the boundary cation relative to the interior cation of the same species, with the color scale in \AA$^3$ and saturated at $\pm 3$~\AA$^3$. The data is from a slice of a relaxed 0~K cell.
The boundary plane is shown by green dashed line, metal atoms of the different grains are orange, purple and green, boron is white. 
(b)~Metal fraction at the boundary sites against temperature. 
The dotted line is the nominal 20~at.\%, error bars are block standard errors.
Gray band shows the region where simulated cell of polycrystalline HEB$_2$ is molten at considered conditions of 3000~K. 
(c)~Segregation energies at 2000~K from the swap spectrum (open circles, error bars from a bootstrap over sites) and from the McLean isotherm (filled squares).
A negative value indicates that the boundary includes the species.
For $\Sigma 7$ tilt the two site classes are shown separately, the compressed $CN_B = 10$ sites (10 boron neighbours instead of 12 in the bulk) (blue, down triangles) and the expanded $CN_B = 14$ sites (red, up triangles), colored as in (a).
McLean values for Hf and Zr on the $CN_B = 10$ sites are absent, since these metals never occupy them.
Open symbols represents the spectrum, and filled symbols represent isotherm.
The columns are divided into chemistry-driven ($\Sigma 6$, $\Sigma 7$ twist, polycrystal, $\Sigma 193$), mixed ($\Sigma 8$) and geometry-driven ($\Sigma 7$ tilt).
The y-axis scales in (c) are different between panels, and tilt values are gauged over Ti, Nb and Ta only (see Supplementary Sec.~\ref{si:md-seg}). 
Swap spectra are provided in Supplementary Fig.~\ref{si:fig-spectrum}.}
\label{fig:gb_main}
\end{figure}

Fig.~\ref{fig:gb_main} collects the MC/MD compositions of six boundaries at 2000--3000~K, the segregation energy at 2000~K and the map of site volume at each boundary. 
The boundaries can be divided into two types based on this map, with $\Sigma 8$ in the middle, and the type determines which metals will segregate.

Where every site at the boundary is larger than in the bulk, segregation is controlled by chemistry.
This is the case for $\Sigma 6$, $\Sigma 7$ twist, $\Sigma 193$ and the general boundaries of the polycrystalline sample: at all of them Nb and Ta are enriched and Hf is depleted, even though the expanded sites would prefer Zr and Hf.
Which of Ti and Zr is also depleted depends on the boundary.
Ti is depleted at $\Sigma 6$ and $\Sigma 193$ (by 12.5 and 12.9~at.\% at 2000~K), while Zr at the twist (by 7.2~at.\%). 
The effect is strongest at the $\Sigma 7$ twist boundary, 35~at.\% of Nb and 35~at.\% of Ta at 2000~K against 7~at.\% each of Zr and Hf. 
With temperature it weakens at the twist (Nb and Ta at 25~at.\% at 3000~K) and at $\Sigma 193$ (from 26 and 30 to 23 and 25~at.\%, respectively for Nb and Ta), while at $\Sigma 6$ it stays at 25--29~at.\% up to 3000~K. 
The $\Sigma 193$ runs started from a random decoration, so this enrichment is reached by the MC. 
The polycrystal has a similar behavior to the ordered grain boundaries with 23 and 24~at.\% of Nb and Ta at 2000~K. 
Its boundaries are less ordered compared to others, and it melts at 3000~K; with 2919 atoms in three grains, the grains are approximately 2~nm in size and about 450 out of 973 cations (46\%) are located at the boundary sites, so the interior acts as a small reservoir.
A larger cell would be beyond the single-GPU limit in Section.~\ref{sec:md}, since the mean site volume is 18\% higher than the bulk value, and Nb and Ta occupy these sites equally, so the enrichment does not require an ordered boundary for its appearance.

The enrichment is shown in the site energies (Fig.~\ref{fig:gb_main}(c); the spectrum is taken over the energy-defined boundary sites in Supplementary Section~\ref{si:md-seg} and the isotherm over the structural ones). 
At all four grain boundaries of this kind, both routes give negative energies for Nb and Ta, ranging from $-0.03$ to $-0.17$~eV from the spectrum and $-0.05$ to $-0.15$~eV from the isotherm.
Positive energies are found for Ti (except for the isotherm at the twist, $-0.004$~eV), which agree within 0.02~eV everywhere except the twist.
Hf and Zr showed different results: the spectrum put them within 0.06~eV of zero (except Zr in the polycrystal, $-0.12$~eV), while the isotherm ranged up to $+0.15$~eV. 
The routes agree in sign for metals that segregated and differed in size by a factor of 2--3, in both directions.
For Nb and Ta the isotherm is the larger of the two at $\Sigma 6$ and the twist, and the smaller at $\Sigma 193$ and in the polycrystal.
The two are not the same in quantity (Section~\ref{sec:md}).
Te spectrum is one exchange in a random boundary at 0~K and the isotherm is an equilibrated boundary at 2000~K.
We take only the sign from them together.

Where the boundary between expanded and compressed sites is located, geometry takes over. 
At the $\Sigma 7$ tilt boundary, there are two sites with an equal number of boron atoms with coordination numbers of 14 and 10, and atomic volumes that are 18\% larger and 13\% smaller than those in the bulk crystal.
The runs fill these sites according to the radius of the segregant.
At 2000~K Zr and Ti accounts for 40 and 39~at.\% of the boundary, respectively, i.e. each occupies most of its own class of sites.
Zr occupies the large site, while Ti occupies the smaller one.
Nb and Ta are depleted to 5--8~at.\% at this grain boundary.
Site energies indicate the same trend, and are an order of magnitude stronger than at any other boundary. 
On the $CN_B = 14$ sites Zr is at $-0.97$~eV and Hf at $-0.70$~eV ($-0.85$ and $-0.59$~eV from the isotherm)
Hf would also occupy the large site, but it is lost to Zr.
On the $CN_B = 10$ sites Ti is at $-0.26$~eV ($-0.22$~eV), and Hf and Zr are 1.4 and 1.7~eV uphill. 
$\Sigma 8$, with a weaker splitting of the site volumes, is a mixed case.
At 2000~K Zr is enriched to 27~at.\% and Ti is depleted to 9~at.\%, as by size with isotherm energies of $-0.07$ and $+0.14$~eV.
Nb and Ta are enriched to 23 and 25~at.\%, as at the chemical kind.
The entire spectrum lies within $\pm 0.03$~eV. 
The effect weakens with temperature at a different rate for each element.
Nb and Ta stop segregating towards $\Sigma 8$ at $T = 2500$~K, while Zr remains at 27-30\% up to 3000~K. 
At $\Sigma 7$, all five metals return to within 7\% of their nominal value at 3000~K.

The tilt site energies can be compared with DFT without new calculations, as the sites of the pentagonal and heptagonal rings of this boundary were calculated for dilute solutes previously.
Re-referencing to Ti, the expanded $CN_B = 14$ site gives $-1.02$, $-0.75$ and $-0.16$~eV for Zr, Hf, and Nb, respectively, compared to $-1.04$, $-0.85$ and $-0.13$~eV at the heptagonal site of TiB$_2$\cite{dai2020strategy}.
Re-referencing to Zr, the compressed $CN_B = 10$ site gives $-1.59$ and $-1.63$~eV for Nb and Ta, respectively, compared to $-1.49$ and $-1.36$~eV at the pentagonal site of ZrB$_2$\cite{dai2017segregation}. 
The hosts differ (dilute solutes in a binary vs HEB$_2$), and the comparison covers the geometric sites where the energies vary by 0.1--1~eV.
The chemical energies, $-0.03$ to $-0.17$~eV, are comparable to the error of the potential on $\gamma_{GB}$ converted to one boundary cation (-0.06 to 0.2 eV), so we can use the potential to determine the sign of Nb and Ta enrichment but not their magnitude.
The potential bias on the binaries (Section~\ref{sec:res-pot}) is working against Nb and Ta.

\begin{figure}[H]
\centering
\includegraphics[width=\textwidth]{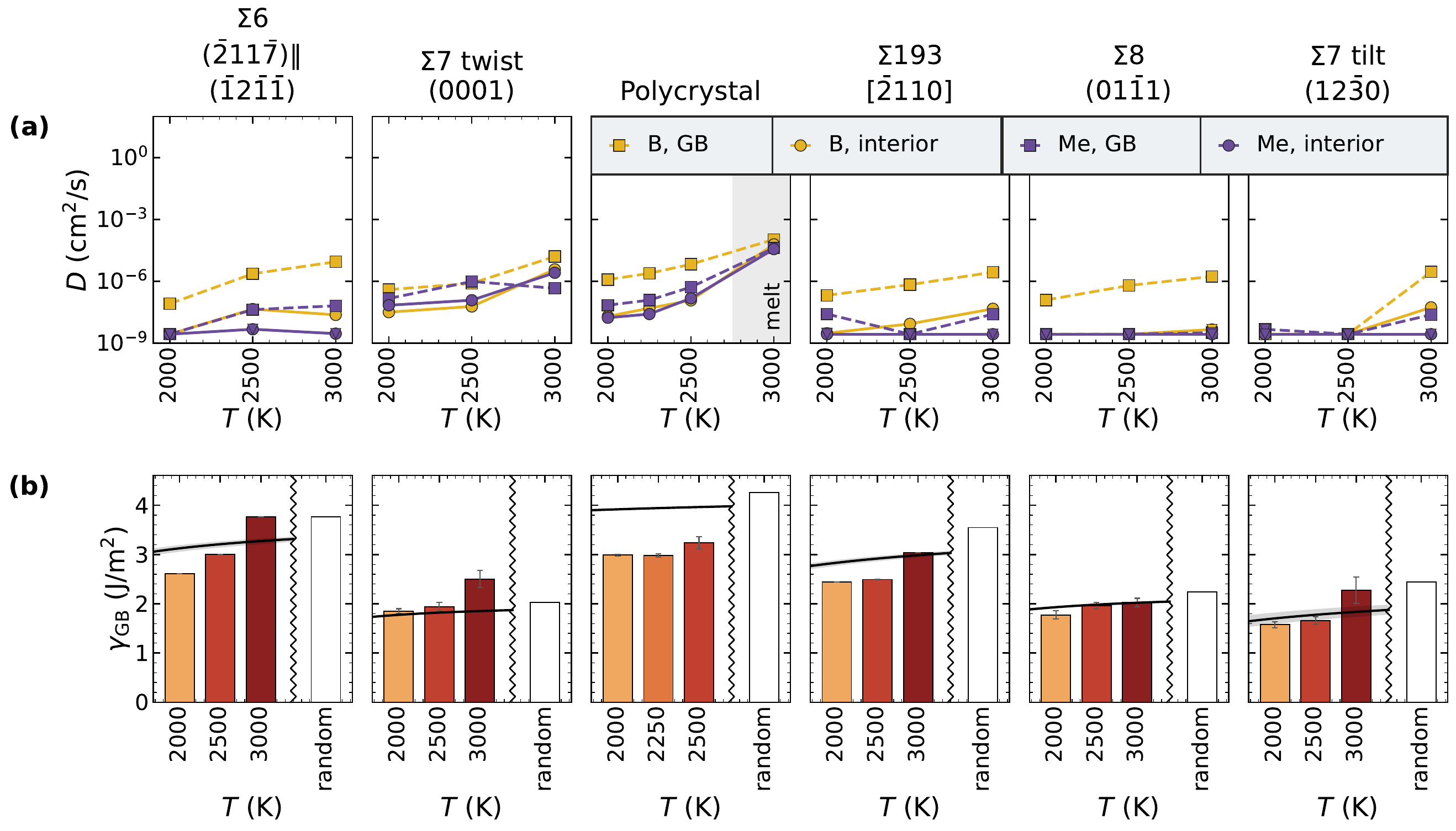}
\caption{Migration of atoms and boundary energy at the six grain boundaries. (a)~Self-diffusion coefficients of boron (yellow) and metal (purple) atoms at the boundary (dashed lines, squares) and in the grain boundary (solid lines, circles).
Points below the resolution of $5\times10^{-9}$~cm$^2$/s are placed on a common floor at the bottom of the axis.
Grey band shows the melting of the polycrystal.
(b)~$\gamma_{GB}$ of MC/MD snapshots quenched to 0~K, by run temperature, and of the relaxed initial cell of the runs, with a random decoration and the boundary geometry as built (white bar, right of the axis break). The black line is the $\gamma_{GB}$ predicted from the swap spectrum, $\gamma_{rand} + \Delta\gamma(T)$, with a bootstrap band over sites. Molten runs are not shown.}
\label{fig:gb_kinetics}
\end{figure}

The loss of segregation with temperature is accompanied by boron diffusion across the sample volume towards the grain boundaries (see Fig.~\ref{fig:gb_kinetics}(a)).
At 2000~K, boron at the boundary diffuses at a rate of $D = 8 \times 10^{-8} - 10^{-6}~\text{cm}^2/\text{s}$ at every boundary, except for the $\Sigma 7$ tilt. 
This is at least an order of magnitude faster than the rate of diffusion in the grain interior (circles on the Fig.~\ref{fig:gb_kinetics}(a)). 
The metals at the boundaries stay below $10^{-7}~\text{cm}^2$/s, and in the interior, mostly below the resolution.
The exception to this is the twist, which has a diffusion coefficient of up to $10^{-6}~\text{cm}^2/$s at 2500~K.
Its two grains slide against each other along the basal plane as rigid blocks. 
The local-sublattice reference in Supplementary Section~\ref{si:md-diff} reduces this motion partially, but at 3000~K the residual sliding also sets the diffusion coefficient of the metal in the twist region.
The cation sublattice, on which the compositions in Fig.~\ref{fig:gb_main}(b) were measured, is static, while boron is the element that rearranges at the boundary. 
At the $\Sigma 7$ tilt boundary, boron diffusion was below the resolution until 2500~K, when it jumped to $2.9\times10^{-6}$ cm$^2$/s at 3000~K. 
This is the same temperature where the size sorting collapsed.
Two factors contribute to this collapse and are not discussed here.
The simple Boltzmann weakening of the fixed segregation energy with $T$, and the growth of the boundary itself. 
As the number of boundary sites increases from 24 at 2000~K to 36 at 3000~K (and the twist changes from 24 to 52), the new sites diffuse the old ones, diluting them. 
Boron diffusion alone cannot reverse segregation: at $\Sigma 8$, boron diffuses at the boundary from 2000~K on, and zirconium remains enriched to 27-30\% up to 3000~K. 
In the polycrystal, at 3000 K, boron and metal both diffuse at 1$10^{-5}$--$10^{-4}$~cm$^2$/s at the boundary and in the grains, which means it is molten.

Fig.~\ref{fig:gb_kinetics}(b) shows the effect of segregation on the boundary energy at different temperatures.
At 2000~K, every boundary has a lower energy than the random case: by 0.17~J/m$^2$ at the twist, 0.47~J/m$^2$ at $\Sigma 8$, and 0.86-1.28~J/m$^2$ at $\Sigma 7$ tilt, $\Sigma 193$, $\Sigma 6$, and in the polycrystal.
With increasing temperature, $\gamma_{GB}$ increases again. 
At 3000~K, it returns to the random value at $\Sigma 6$ (3.77 vs 3.76~J/m$^2$) and is above it for the twist (2.50 vs 2.02~J/m$^2$), but it remains 0.17-0.51~J/m$^2$ lower than the random value for the tilt, $\Sigma 8$ and $\Sigma 193$ (2.27 vs 2.44~J/m$^2$, 2.02 vs 2.24~J/m$^2$, and 3.04 vs 3.55~J/m$^2$ respectively).
The twist still has 25\% Nb and Ta at 3000~K, so its excess energy over the random value is not due to chemistry.
The quench retains some of the thermal disorder in the boundary, which is not separated from the chemistry in the other samples either.
The prediction from the swap spectrum accurately reproduces the twist and $\Sigma 7$ tilt at 2000 and 2500~K within a range of 0.07-0.16~J/m², which is comparable to the spread of the snapshots.
Additionally, $\Sigma 8$ is also accurately predicted at 3000~K.

At these three boundaries, the drop of $\gamma_{GB}$ is determined by the site preferences at the fixed 0 K boundary.
At 3000 K, the tilt and twist lie above the prediction based on thermal disorder, which is kept by the quench.
At $\Sigma 6$ and $\Sigma 193$, the measured drop at 2000 K is larger than predicted by 0.5 and 0.4 J/m$^2$, respectively, and in the polycrystal, it is 0.9 J/m$^2$ larger.
The prediction accounts only for the chemical preferences of the sites at a fixed boundary. 
We attribute the rest to the reconstruction of these boundaries, mainly their boron network, during the process, which cannot be captured by the spectrum, since boron never swaps and relaxes only locally around an exchange.
MC also swaps cations in grain interiors, and any short-range order they develop is counted as boundary energy, because the bulk reference is random.
The boundaries that the prediction produces indicate that this area is small: its interior is similar in nature to that of the other cells. 
In this region, the prediction, which does not include an interior term, leaves a maximum of 0.16~J/m$^2$ unexplained.
For the polycrystal, this can be verified.
Its 2500 K geometry with the cation positions randomized (not shown in Fig. \ref{fig:gb_kinetics} (b)) has a GB energy of 3.42 J/m$^2$, so when the geometry is annealed alone, it reduces the as-built GB energy by 0.85 J/m$^2$. 
From this reference, the predicted value (3.08 J/m$^2$ at 2000 K) matches the measured value (2.99 J/m$^2$) within 0.1 J/m$^2$. 
Therefore, roughly two thirds of the reduction is due to reconstruction, and one third is due to chemistry. 
The reference geometry is at 2500 K, so the proportion is valid to the extent that the GB boundary at 2000 and 2500 K is the same.
For ordered boundaries, the same randomization was not applied.

The two types of boundaries, with $\Sigma 8$ in between, separate different metals.
The type of boundary depends on its site volume map, which can be ordered or general. 
For a polycrystal, the chemical type is most important, as general boundaries belong to this category.
The segregation of Nb and Ta agrees in sign with the density functional theory (DFT) energetics of ordered boundaries.
The $\gamma_{GB}$ value of the group V diboride is lower than that of group IV, so a boundary enriched with Nb and Ta is cheaper. 
However, this agreement is only qualitative.
When converted to energy per boundary cation, the difference between the binary duborides is several times greater than the segregation energies in Fig.~\ref{fig:gb_main}c. 
It is discussed in Section~\ref{sec:gb-dft}.
The role of the extra valence electrons of Nb and Ta in the bonding is discussed in Section~\ref{sec:cohp}.

The geometric kind is what was previously found in diborides, in ZrB$_2$ with Nb and Ta taking the compressed boundary sites\cite{dai2017segregation}, and in (Ti,Zr,Hf)B$_2$, the smallest cation, Ti, segregates\cite{dai2024deep}. 
Both of these are in line with the size.
Chemical species such as Nb and Ta occupy expanded sites, but their size does not predict this.
A binary boundary that is cheaper does not always determine segregation in a lattice.
In the high-entropy carbide, with the same five metals, a Monte Carlo simulation with a universal MACE model (not fine-tuned) at 300 and 2000 K puts Ti and Zr on the $\Sigma 5$ boundary \cite{mou2026finite}. 
Although there too, the group V binary species have the cheaper grain boundaries. 
The values for TiC, ZrC, and HfC are between 1.88 and 2.96~J/m$^2$ in DFT\cite{mou2026finite}, while the values for NbC and TaC are 2.38 and 2.69~J/m$^2$. 
In the diboride, the cheaper boundaries for Nb and Ta coincide with the change in bonding at $E_F$, as discussed in Section~\ref{sec:cohp}.
We did not check whether the carbide lacked this.

The measurement of host metal at the HEB$_2$ boundary does not show enrichment. 
STEM-EDS maps across the boundary of HEB$_2$, created by reactive spark plasma sintering with a final 10-minute hold at 2000~$^\circ$C (2273 K), show all five metals to be homogeneous \cite{wang2021synergic}.
The maps shown are qualitative and do not provide a profile of the composition at a given point.
It is not known whether the sample has reached equilibrium, as in the case of MC samples. 
A profile of the composition across the boundaries of an annealed sample would provide a way to test the prediction.

\subsection{COHP analysis of bonding at the grain boundaries}
\label{sec:cohp}

\begin{figure}[H]
\centering
\includegraphics[width=\textwidth]{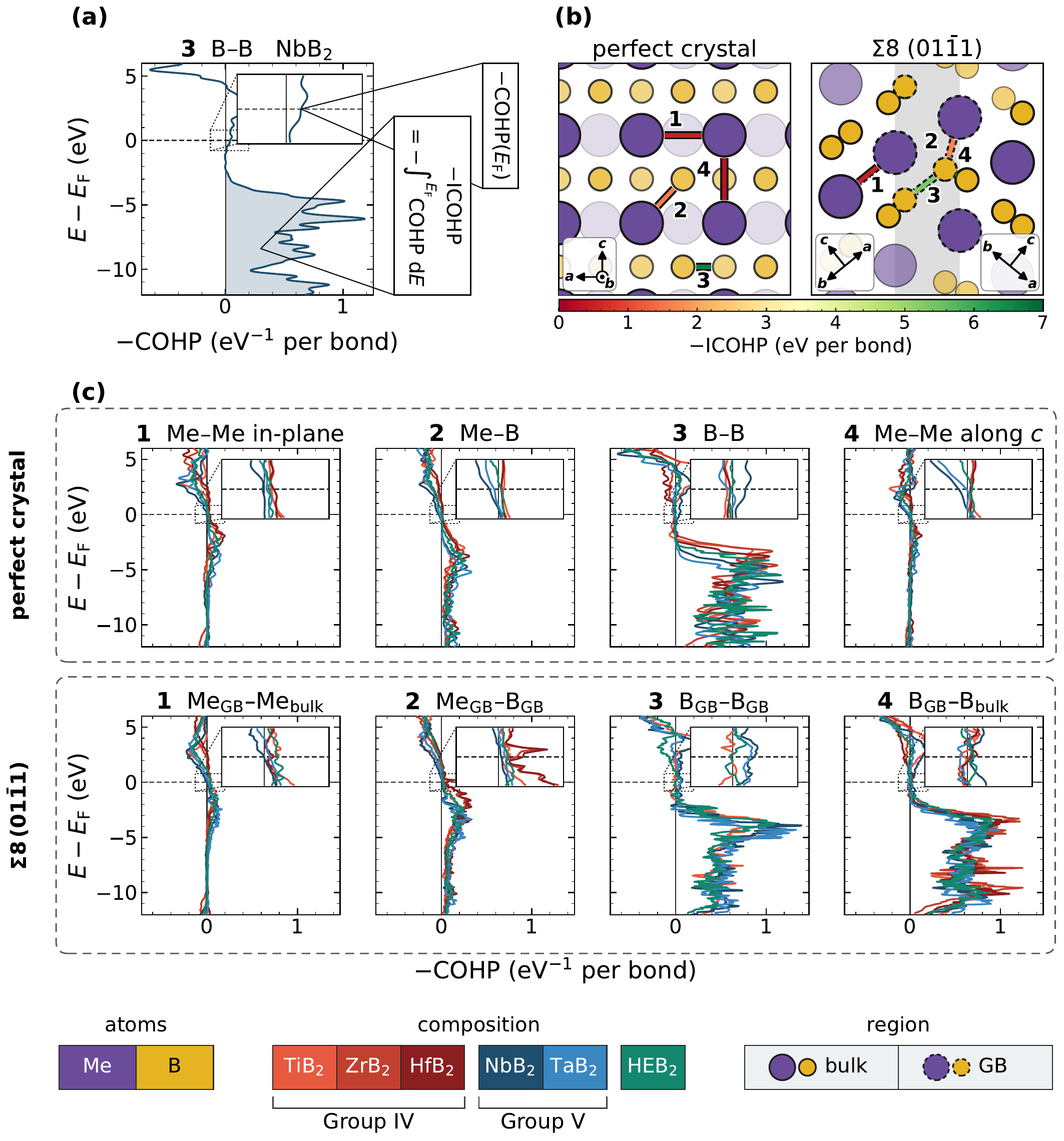}
\caption{COHP curves of the bonds in the bulk and at the grain boundary. (a)~Example: the B--B bond in bulk NbB$_2$; the shaded area up to $E_F$ is $-$ICOHP, the inset shows $-$COHP$(E_F)$, a positive value means bonding states at the Fermi level. (b)~The four bond types of the bulk crystal (left) and the four bond classes at the $\Sigma 8$~$(01\bar{1}1)$ boundary (right, boundary atoms outlined with dashes, boundary slab in grey) for TiB$_2$; the color of a bond is its $-$ICOHP. (c)~$-$COHP against energy for each bond type, upper row bulk, lower row $\Sigma 8$; the curve of a composition is the mean over all bonds of that class, insets magnify the window $E - E_F = \pm 0.8$~eV. Atom colors of (b) and composition colors of (a, c) as in the two keys.}
\label{fig:cohp}
\end{figure}

To analyze chemical bonding in grain boundaries in HEB$_2$ two types of descriptors are used, shown in Fig.~\ref{fig:cohp}(a) for the example of B--B bond in NbB$_2$.
These descriptors are $-$COHP$(E_F)$, whose sign tells whether the states at the Fermi level are bonding or antibonding, and $-$ICOHP, the integral up to $E_F$, i.e. the total strength of the bond. 
For the bulk four types of bonds can be considered, see Fig.~\ref{fig:cohp}(b), namely Me--Me in-plane (denoted as 1 in Fig.~\ref{fig:cohp}(b)), Me--B (2), B--B (3) and Me--Me inter-plane (4). 
At the grain boundary $\Sigma 8$ also four types of bonds are considered, as shown in right panel of Fig.~\ref{fig:cohp}(b), namely Me$_{GB}$--Me$_{bulk}$ (1), Me$_{GB}$--B$_{GB}$ (2), B$_{GB}$--B$_{GB}$ (3) and B$_{GB}$--B$_{bulk}$ (4).
At the base of the twist, the last layer is empty, as the next layer along the $c$-axis lies outside the first layer, and the bonds between the Me$_{GB}$ and Me$_{bulk}$ atoms are along the $c$-direction.

The COHP curves of all compositions are given in Fig.~\ref{fig:cohp}(c) for $\Sigma 8$ and for all four boundaries in Supplementary Fig.~\ref{si:fig-cohp-per-gb}; the descriptors, averaged over $\Sigma 6$, $\Sigma 7$ tilt, $\Sigma 7$ twist and $\Sigma 8$, are collected in Fig.~\ref{fig:icohp}, and per boundary in Supplementary Table~\ref{si:tab-cohp}.

By ICOHP the bonds at the boundary hardly differ from the bulk ones.
Me--B bonds at the boundary are stronger than in the bulk by 0.10--0.27~eV (from 1.59--1.87~eV in the bulk to 1.70--2.07~eV at the boundary), Me--Me bonds by 0.01--0.13~eV, for all six compositions.
B$_{GB}$--B$_{bulk}$ bonds are as strong as the bulk ones (5.8--6.6 against 5.8--6.7~eV), and the mean of the B$_{GB}$--B$_{GB}$ bonds moves by $-0.17$ to $+0.09$~eV, with a tail of a few bonds with $-$ICOHP down to 2.9~eV at $\Sigma 6$ and the $\Sigma 7$ tilt (see violins in Fig.~\ref{fig:icohp}(c)).
All these shifts are smaller than the spread of $-$ICOHP over the bonds of one class (standard deviation 0.4--0.9~eV).

\begin{figure}[H]
\centering
\includegraphics[width=\textwidth]{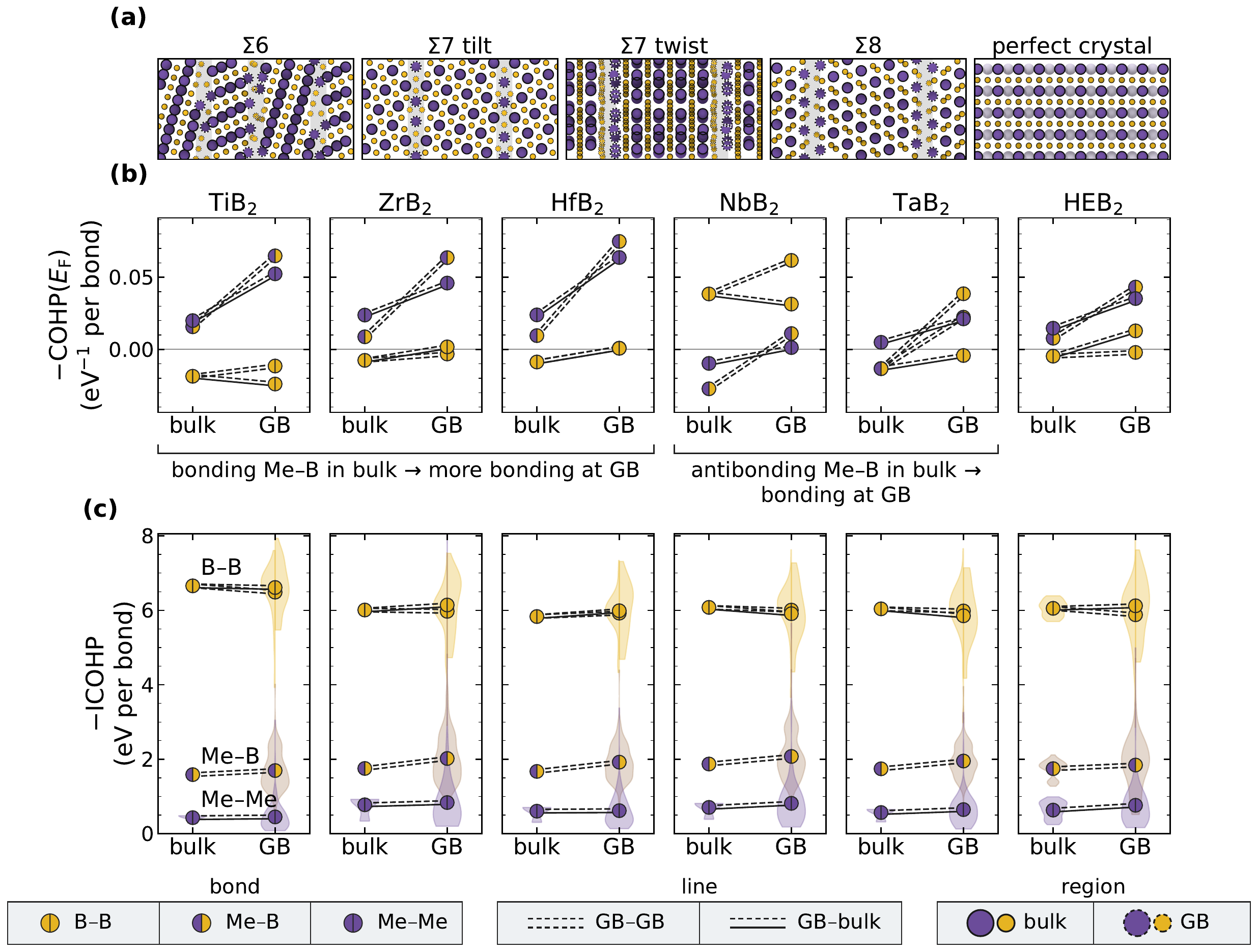}
\caption{Bond descriptors on going from the bulk to the grain boundary. (a)~The $\Sigma 6$, $\Sigma 7$ tilt, $\Sigma 7$ twist and $\Sigma 8$ cells and the bulk crystal.
Boundary atoms are marked with dashed rings, boundary region in grey. 
(b)~$-$COHP$(E_F)$ and (c)~$-$ICOHP of the B--B, Me--B and Me--Me bonds in the bulk and at the boundary for the five binary diborides and HEB$_2$; the boundary point is the mean over all bonds of that class at the four boundaries (the twist, with the largest boundary area, contributes most of them and has no B$_{GB}$--B$_{bulk}$ bonds), the violins in~(c) show the distribution over individual bonds. 
The two halves of a marker are colored by the two atoms of the bond (metal purple, boron yellow), so B--B, Me--B and Me--Me read from the marker.
The three types of bonding are labeled explicitly for the TiB$_2$ panel of~(c). 
Double dashed lines join the bulk bond to the boundary--boundary bonds, dashed-solid lines are referred to the boundary--bulk bonds. 
Per-boundary values are presented in Supplementary Table~\ref{si:tab-cohp}.}
\label{fig:icohp}
\end{figure}

The difference between groups IV and V can be observed using $-$COHP at the Fermi level.
In the bulk TiB$_2$, ZrB$_2$ and HfB$_2$ the states at $E_F$ are bonding for both Me--B ($-$COHP$(E_F)$ = 0.009--0.016~eV$^{-1}$) and for Me--Me (0.020--0.024~eV$^{-1}$).
In contrast, in NbB$_2$ and TaB$_2$ the Me--B bonds are antibonding at $E_F$ with values of $-0.027$ and $-0.013$~eV$^{-1}$, respectively. 
The Me--Me bond is antibonding for Nb--Nb ($-0.010$~eV$^{-1}$) and close to zero for Ta--Ta ($+0.005$~eV$^{-1}$). 
For the B--B bond, the situation is reversed.
Only NbB$_2$ has bonding states at $E_F$ with $-$COHP$(E_F)=+0.038$~eV$^{-1}$, as shown in Fig.~\ref{fig:cohp}(a)).
The other five compositions have weakly antibonding states ($-0.005$ to $-0.019$~eV$^{-1}$). 
Overall, it can be seen that the Fermi level of group V is located in antibonding Me--B states in the bulk.
This means that the extra electron in Nb and Ta occupies into states that weaken the Me--B bond.

At the grain boundary the signs change.
For group IV, the Me--B bonds at the boundary become 4--8 times stronger, where $E_F$ is ranging from 0.064 to 0.075~eV$^{-1}$ and the Me--Me bonds become 2--3 times stronger.
For group V, the antibonding states disappear: the Me--B goes from $-0.027$ to $+0.011$ for Nb and from $-0.013$ to $+0.022$~eV$^{-1}$ for Ta, and the Me--Me goes from $-0.010$ to zero ($+0.001$) for Nb and from $+0.005$ to $+0.021$~eV$^{-1}$ for Ta.
The B$_{GB}$--B$_{GB}$ bonds of group V at the grain boundary are also stronger than B--B in the bulk with $E_F$ of $+0.062$ against $+0.038$ for Nb, $+0.039$ against $-0.013$~eV$^{-1}$ for Ta, while the bonds fo group IV are between $-0.012$ and $+0.001$~eV$^{-1}$.
The size of this difference depends on the grain boundary, and roughly follows the change in $\gamma_{GB}$. 
The shift of the group V Me--B states is largest at the twist ($+0.050$~eV$^{-1}$, mean of Nb and Ta), followed by $\Sigma 6$ and $\Sigma 8$ ($+0.039$ and $+0.035$), and smallest at $\Sigma 7$ tilt ($+0.010$~eV$^{-1}$ with Nb--B remaining antibonding at $-0.018$~eV$^{-1}$.
The ratio of $\gamma_{GB}^{group V}$ to $\gamma_{GB}^{group IV}$ from Table~\ref{tab:gb_dft} runs 0.39, 0.54, 0.41 and 0.87 in the same order. 
This trend holds for the two ends (twist and tilt), but for $\Sigma 6$ and $\Sigma 8$ they are reversed, so this is only a general trend with four grain boundaries.
The tilt has the smallest swing and no group splitting, and is also the boundary where segregation in Sec.~\ref{sec:mlip-gb} is determined by geometry.
For HEB$_2$ the descriptors averaged across all metals behave like a group IV (Me--B range from 0.008 to 0.043~eV$^{-1}$), and we did not differentiate them by metal species weithin HEB$_2$.

In short, ICOHP shows that the bonds at the grain boundary are as strong as they are in the bulk for all six compositions.
The groups differ at the Fermi level, where in the bulk for group V diborides $E_F$ corresponds to the antibonding Me--B (and Me--Me for Nb) states.
At the grain boundary these states either leave the Fermi level or become bonding states.
For group IV there are no antibonding states to remove.
The descriptor is read based on its sign, rather than its magnitude.
Group IV gains more bonding at $E_F$ at the boundary compared to group V, and NbB$_2$ is more antibonding in the bulk than TaB$_2$, but its boundaries are more expensive ($\gamma_{GB} (NbB_2) > \gamma_{GB} (TaB_2)$).
So, the argument relies on the antibonding states in group V in the bulk that are lost at the boundary.
Between boundaries, the magnitude of the shift of group V follows how much cheaper its boundaries are in DFT (Section~\ref{sec:gb-dft}).
The same metals, Nb and Ta, segregate chemically in HEB$_2$ (Section~\ref{sec:mlip-gb}). 
This is a correlation between Fermi-level descriptor and the boundary energy.
COHP$(E_F)$ is not an energy and we do not derive $\gamma_{GB}$ from it.

\subsection{Strength of HEB$_2$ and of a segregated boundary}
\label{sec:strength}

\begin{figure}[H]
\centering
\includegraphics[width=\textwidth]{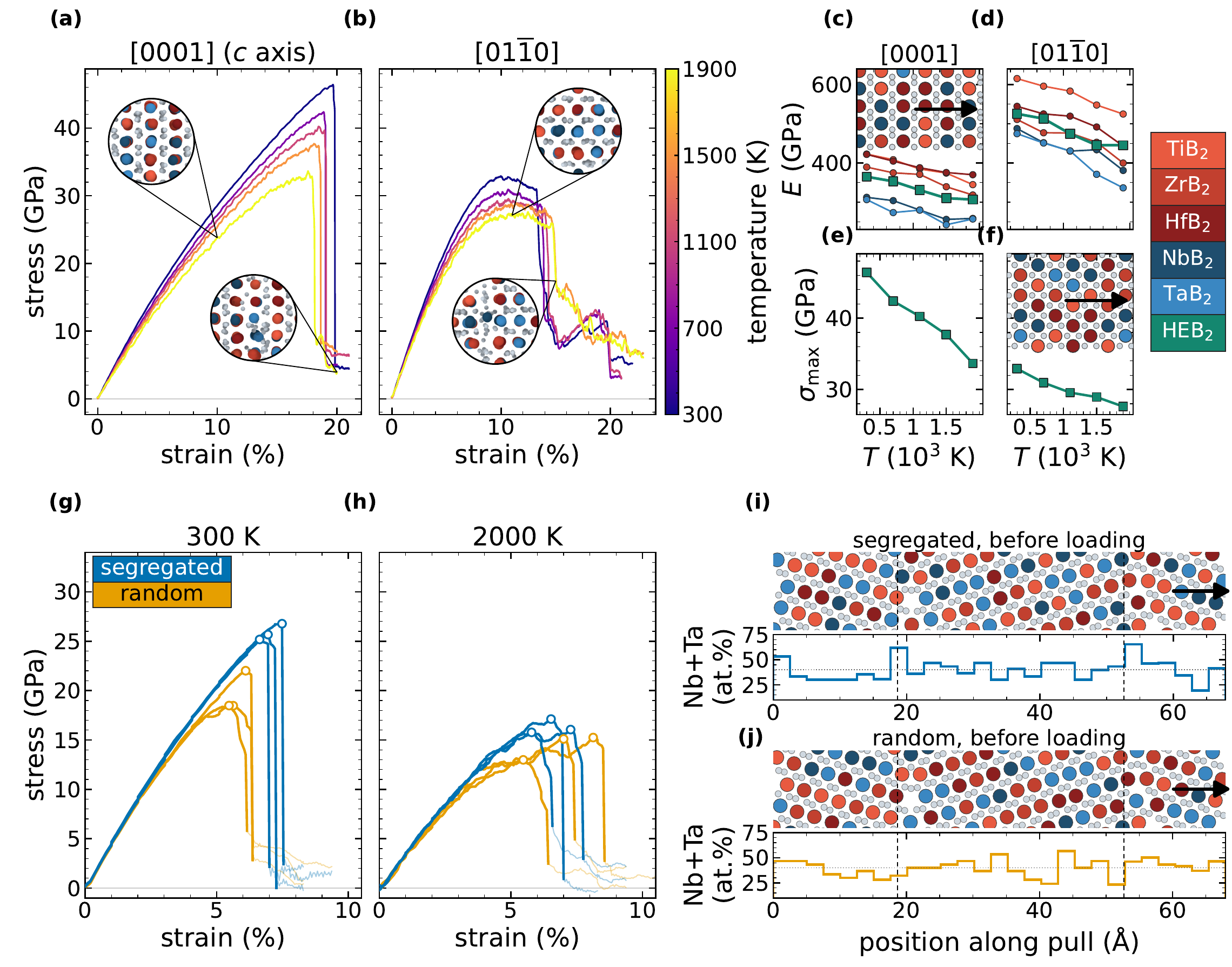}
\caption{Uniaxial tension of single-crystal and $\Sigma 193$ boundary of HEB$_2$ with MACE. For single crystal.
(a,b)~Averaged stress--strain curves along [0001] and $[01\bar{1}0]$ directions at 300--1900~K (see color scale).
The circular insets represent slices of the cell at the strains they point to, before and after failure. 
(c,d)~Young's modulus $E$ against temperature for HEB$_2$ (teal) and the five binary diborides (colors as in the key), from stepped finite deformation.
The cell sketches show the direction of tensile deformation. 
(e,f)~Maximum stress $\sigma_{max}$ of HEB$_2$ against temperature. 
Averaged stress--strain curves of three segregated cells (blue; cation placements taken from 2000~K runs) and three random cells (orange; the same cells with the cations permuted at random) calculated at (g)~300 and (h)~2000~K.
Open circles mark $\sigma_{max}$ for each run. (i, j)~Atomic structures of segregated and random cells before loading at 300~K. Below each structure the Nb+Ta fraction of all cations in the cell is plotted in 2.5~\AA{} slices, with a nominal 40~at.\% dotted line.
Dashed lines mark the two boundary planes.}
\label{fig:tensile_mono}
\end{figure}

We first performed tensile deformation on the HEB$_2$ single crystal (Fig.~\ref{fig:tensile_mono}(a,b)). 
Along the [0001] direction the ultimate tensile strength (UTS) is 46.4~GPa at 300~K and sample fails at a strain of 20\%.
Along the $[01\bar{1}0]$ direction the UTS is lower, 32.9~GPa at a strain of 10\%, followed by a plateau of a few percent before failure.
As temperature increases both values decrease: $\sigma_{max}$ along the [0001] decreases to 33.7~GPa and along the $[01\bar{1}0]$ it decreases to 27.6~GPa at 1900~K.
The Young's modulus of HEB$_2$ is the average of the five binary diborides values (Fig.~\ref{fig:tensile_mono}(c,d)): 365 against 370~GPa along the [0001] and 525 against 527~GPa along the $[01\bar{1}0]$ at 300~K. 
Along the [0001] it stays within 3\% of this average p to 1900~K, along the $[01\bar{1}0]$ it stays within 3\% up to 1500~K and 7\% above it at 1900~K (445 against 417~GPa). 
The stiffness along the $c$-axis is also where the potential accuracy is lowest ($C_{33}$ is within 3\%, Table~\ref{tab:mlip-benchmark}b), so the values along the [0001] direction carry this uncertainty.

% \begin{figure}[H]
% \centering
% \includegraphics[width=\textwidth]{Figs/fig_tensile_gb.pdf}
% \caption{Tension of the $\Sigma 193$ boundary of HEB$_2$ along its normal. 
% Averaged stress--strain curves of three segregated cells (blue; 0~K quenches of the 2000~K MC/MD run) and three random cells (orange; the same cells with the cations permuted at random) calculated at (a)~300 and (b)~2000~K.
% Open circles mark $\sigma_{max}$ for each run, and the part of a curve after the failure is drawn faintly. 
% Atomic structures of (c)~Segregated and (d)~random cells before loading at 300~K. 
% A 3.7~\AA{} slab viewed along the boundary planes, metals are colored in the same way as in Fig.~\ref{fig:tensile_mono}, boron atoms are small and grey, the arrow determines the  direction of tensile deformation.
% Below each structure the Nb+Ta fraction of all cations in the cell is plotted in 2.5~\AA{} slices, with a nominal 40~at.\% dotted line.
% Dashed lines mark the two boundary planes.
% The maximum stress $\sigma_{max}$ (mean $\pm$ standard deviation over three cells) is $25.9\pm0.8$ against $19.7\pm2.0$~GPa at 300~K and $16.3\pm0.7$ against $14.4\pm1.3$~GPa at 2000~K for the segregated and the random cells, respectively.
% $E$ is $451\pm6$ against $435\pm3$~GPa and $343\pm2$ against $324\pm10$~GPa.}
% \label{fig:tensile_gb}
% \end{figure}

The tensile deformation along the normal of the segregated and random $\Sigma 193$ grain boundary of HEB$_2$ is shown in the Fig.~\ref{fig:tensile_mono})(g,h,i,j).
As can be seen the $\Sigma 193$ boundary fails at 20--26~GPa (group means) and 6--7\% strain at 300~K, against 33 and 46~GPa at the strain of 10 and 20\% for the single crystal along two directions, and in every run the crack opens at a boundary (Supplementary Fig.~\ref{si:fig-crack}). 
Before loading (Fig.~\ref{fig:tensile_mono}(i,j)) the two boundary planes of the segregated cell hold 62 and 65~at.\% of Nb and Ta against about 37~at.\% in the grains, while in the random cell the same planes hold 32 and 46~at.\%, within its fluctuations from slice to slice. Segregation makes this boundary stronger: at 300~K the three segregated cells hold $25.9\pm0.8$~GPa and the three random ones $19.7\pm2.0$~GPa, at a strain of 7.0 against 5.7\%, and the segregated cells are also 4\% stiffer. At 2000~K the difference shrinks to $16.3\pm0.7$ against $14.4\pm1.3$~GPa; every segregated cell is still above every random one, but the difference is comparable to the spread of the random cells. Since within each pair of cells the geometry is the same and only the cations are rearranged, the gain at 300~K comes from the chemistry of the boundary, with the caveat that this common geometry was annealed together with the segregated chemistry. It goes in the direction expected from the boundary energy: at fixed surface energy the work of separation is $W_{sep} = 2\gamma_s - \gamma_{GB}$, and segregation lowers $\gamma_{GB}$ of $\Sigma 193$ by 1.1~J/m$^2$ against its random reference of Fig.~\ref{fig:gb_kinetics}(b) ($\gamma_{GB}$ of the tension cells themselves was not computed). The surface term was not computed either, and it works the other way: in DFT the Nb- and Ta-rich surfaces are the cheaper ones (Supplementary Table~\ref{si:tab-surf}), so a Nb- and Ta-rich fracture surface lowers $W_{sep}$. The three segregated cells are snapshots of one MC/MD run, 20~ps apart, so this is one boundary and one run per chemistry, and the result shows that segregation of Nb and Ta can strengthen a boundary. For Nb and Ta, earlier DFT calculations give either sign, depending on the host: on the sites they prefer they raise the ideal tensile strength of ZrB$_2$ boundaries by 12.5--34\%\cite{dai2017segregation} and lower it at all three TiB$_2$ boundaries studied, by up to 24\%\cite{dai2020strategy}. W, a different segregant, strengthens 13 of 14 boundaries of ZrB$_2$, by up to 45\% in MD at 2000~K\cite{dai2020grain}. Our gain of 31\% at 300~K is of the size found for ZrB$_2$, for a segregant that makes up most of the boundary plane rather than a single site.

\FloatBarrier
\section{Discussion}
According to our calculations $\gamma_{GB}$ splits by group at the $\Sigma 6$, $\Sigma 8$ and basal twist boundaries with grain boundary energies of 2.5--4.8~J/m$^2$ for group IV, 0.8--2.7~J/m$^2$ for group V, with TaB$_2$ having the lowest energy everywhere and NbB$_2$ with highest energy. 
The symmetric $\Sigma 7$ tilt boundary shows no split, and for TiB$_2$ and ZrB$_2$ it agrees with earlier DFT within 0.06~J/m$^2$. 
A random decoration of HEB$_2$ gives average value of $\gamma_{GB}$ among the binary diborides, within 0.1~J/m$^2$ at three grain boundaries and 0.23~J/m$^2$ above the average at $\Sigma 8$, and the large considered cell with MACE agree with the average value within 0.12~J/m$^2$.

By ICOHP analysis, the Me--B and Me--Me bonds at the boundary are, on average, as strong as or slightly stronger than in the bulk, by 0.01--0.27~eV per bond. 
The difference between the two groups is at the Fermi level. 
In the bulk of NbB$_2$ and TaB$_2$, the Me--B bonds are antibonding at $E_F$ (and so are Nb--Nb bonds, while Ta--Ta is close to zero, $+0.005$~eV$^{-1}$).
However, directly at the boundary, they become bonding, on average over four boundaries.
For group IV, Nb--Nb and Ta--Ta bonds are bonding in both cases. 
The shift is largest at the twist boundary and smallest at the $\Sigma 7$ tilt boundary, which has no group split. 
The results for group V are consistent with these changes at $E_F$, and there is a correlation between the four boundaries.

In HEB$_2$, the segregation follows a similar pattern. 
At the $\Sigma 6$, $\Sigma 7$ twist, $\Sigma 193$ and the general boundaries of a three-grain polycrystal, both Nb and Ta are enriched to 23--35\% at 2000~K, while Hf is depleted.
This is in contrast to what would be predicted based on size alone. 
Both the swap spectrum and McLean isotherms give negative segregation energies for Nb and Ta at these sites, ranging from $-0.03$ to $-0.17$~eV.
At this scale, the potential determines the sign of the segregation rather than its magnitude.
Only the $\Sigma 7$ tilt, with compressed and expanded sites next to each other, sorts the metals according to their radius: Ti occupies the small site, Zr the large site, and their site energies are within 0.3~eV of those predicted by DFT for dilute solutes in the same positions.
This sorting is lost at 3000~K, along with the onset of boron diffusion at the boundaries. $\Sigma 8$ is in between, where Zr is enriched alongside Nb and Ta at 2000~K and Ti is depleted.
At 2000~K, the segregated boundaries have a separation of 0.17--1.28~J/m$^2$ below the random decoration in $\gamma_{\mathrm{GB}}$.
The site preferences alone account for 0.25--0.74~J/m$^2$, reproducing the drop at twist, $\Sigma 8$ and $\Sigma 7$ tilt; at $\Sigma 6$ and $\Sigma 193$, the rest is attributed to boundary reconstruction during the run.
In polycrystals, two-thirds of the drop comes from the annealing of boundary geometry, and one-third from chemistry. 
At 300~K tension, the $\Sigma 193$ boundary fails at 20--26~GPa, compared to 33--46~GPa for single crystals, whose Young's modulus is the average of five binary values. 
The segregated boundary has a strength of $25.9\pm0.8$~GPa, compared to $19.7\pm2.0$~GPa for a random boundary.

For the ceramic, this has consequences that we did not expect. 
A boundary that is made cheaper by segregation provides less driving force for grain growth, and even the random HEB$_2$ boundaries, on average, are below those of ZrB$_2$ at all four boundaries (by 0.02--0.64~J/m$^2$), which explains the slower grain growth of HEB$_2$ compared to ZrB$_2$\cite{cheng2023grain}. 
Since the hardness of high-entropy diborides depends on the grain size\cite{yang2024effect, feng2021effect}, both a finer grain size and a stronger boundary will affect it. To determine whether they do, we need to measure grain growth and hardness on annealed samples.

Oxidation is outside the scope of this work, but it is important to note that it has an impact on the segregation process in two opposing ways.
In HEB$_2$, the metals oxidize in the following order: Hf, Zr $>$ Ti $>$ Ta, Nb. 
At temperatures between 1500 and 1800~$^\circ$C, Nb and Ta remain in borides, while Hf and Zr have already formed oxides, resulting in a boundary that is enriched in Nb and Ta but depleted in Hf. 
This boundary may take longer to oxidize compared to a random boundary.

Once Nb and Ta have been oxidized, they form eutectic compounds with TiO$_2$ and ZrO$_2$ that melt at temperatures between 1436 and 1621~$^\circ$C.
In the high-entropy carbide, these liquids leave behind voids that allow oxygen to enter through the boundaries. 
The outcome of these processes depends on the conditions of the oxidation run and the potential extension to the oxide phase.

The potential compresses the $\gamma_{GB}$ range by 0.2--0.45~J/m$^2$ at its low end, so it works against Nb and Ta, and it is unlikely that it has produced their enrichment.
DFT cells contain one decoration per boundary, and SQS cells do not place Nb on a boundary site.
Therefore, the bonding of Nb and Ta inside HEB$_2$ is inferred from binaries. 
The electron count argument is specific to diborides: in high-entropy carbides of the same metals, group V binaries also have cheaper boundaries, but Ti and Zr still segregate. One experiment on host metals at a HEB$_2$ boundary, a qualitative STEM-EDS map of a spark plasma-sintered sample, showed no enrichment.
A composition profile across the boundaries of an annealed sample would confirm or refute this prediction.

\FloatBarrier
\section{Conclusions}

In this work, we combined density functional theory, crystal orbital Hamilton population analysis, and machine-learning-accelerated Monte Carlo/molecular dynamics to elucidate the grain-boundary behavior of the high-entropy diboride (Hf,Ta,Ti,Nb,Zr)B$_2$ and its binary constituents.
Our calculations reveal a clear group-dependent split in grain-boundary energies: group V diborides (NbB$_2$, TaB$_2$) exhibit roughly half the boundary energy of group IV diborides (TiB$_2$, ZrB$_2$, HfB$_2$) at $\Sigma 6$, $\Sigma 8$, and basal twist boundaries, while the symmetric $\Sigma 7$ tilt remains insensitive to composition.
This distinction originates at the Fermi level, where the Me--B bonds in bulk NbB$_2$ and TaB$_2$ are antibonding but become bonding at the boundary, a change absent in group IV diborides.
In the high-entropy diboride, this electronic effect drives the segregation of Nb and Ta to five of the six studied boundaries, reaching 23--35 at.\% at 2000~K, contrary to size-based predictions.
Only the $\Sigma 7$ tilt, which presents compressed and expanded sites, sorts metals by atomic radius.
Segregation lowers the grain-boundary energy by 0.17--1.28~J/m$^2$ and strengthens the $\Sigma 193$ boundary, which sustains $25.9\pm0.8$~GPa in tension at 300~K compared to $19.7\pm2.0$~GPa for a random boundary.

These findings have implications for the processing and properties of high-entropy diboride ceramics.
The reduced boundary energy due to segregation suggests a lower driving force for grain growth, consistent with the slower grain growth observed in HEB$_2$ relative to ZrB$_2$.
Since hardness in these materials follows a Hall--Petch relationship, a finer grain size combined with stronger boundaries could enhance mechanical performance.
However, the effect of segregation on oxidation resistance remains ambiguous: while Nb- and Ta-enriched boundaries may delay oxidation, they can also form low-melting eutectics with TiO$_2$ and ZrO$_2$, potentially compromising integrity at elevated temperatures.

\vspace{10mm}

\section*{Declaration of generative AI and AI-assisted technologies in the writing process}

During the preparation of this work the authors used Claude Opus (Anthropic) to polish the text of the manuscript and help with code for calculations analysis. All results were rechecked by the authors, who take full responsibility for the content of the publication.

\section*{Data availability}

The MACE potential, its training and validation sets and the atomic structures of this work are available at Zenodo (DOI: 10.5281/zenodo.23453025).


\bibliographystyle{elsarticle-num}

\bibliography{refs}
\end{document}